\documentclass[fleqn,usenatbib]{mnras}

\usepackage{newtxtext,newtxmath,pdflscape}
\usepackage{changepage}
\newcommand{\kk}{K_\mathrm{s}}

\usepackage[T1]{fontenc}

\DeclareRobustCommand{\VAN}[3]{#2}
\let\VANthebibliography\thebibliography
\def\thebibliography{\DeclareRobustCommand{\VAN}[3]{##3}\VANthebibliography}

\usepackage{graphicx}	
\usepackage{amsmath}	

\newcommand{\vrad}{$v_{R'}$}
\newcommand{\vtan}{$v_\phi$}
\newcommand{\srad}{$\sigma_{R'}$}
\newcommand{\stan}{$\sigma_\phi$}
\newcommand{\masyr}{mas yr$^{-1}$}
\newcommand{\rkin}{$R'$}
\newcommand{\kms}{km s$^{-1}$}

\title[VMC XLVIII]{The VMC Survey -- XLVIII. Classical Cepheids unveil the 3D geometry of the LMC}

\author[V. Ripepi et al.]{\noindent 
Vincenzo Ripepi,$^{1}$\thanks{E-mail: vincenzo.ripepi@inaf.it}
Laurent Chemin,$^{2}$
Roberto Molinaro,$^{1}$
Maria-Rosa L. Cioni,$^{3}$
\newauthor
Kenji Bekki,$^{4}$
Gisella Clementini,$^{5}$
Richard de Grijs,$^{6,7}$
Giulia De Somma,$^{1,8}$
\newauthor
Dalal El Youssoufi,$^{3}$
L\'eo Girardi,$^{9}$ 
Martin A.T. Groenewegen,$^{10}$
Valentin Ivanov,$^{11,12}$
\newauthor
Marcella Marconi,$^{1}$
Paul J. McMillan,$^{13}$
Jacco Th. van Loon$^{14}$
\\
$^{1}$INAF-Osservatorio Astronomico di Capodimonte, Salita Moiariello 16, I-80131, Naples, Italy\\
$^{2}$ Centro de Astronom\'ia – CITEVA, Universidad de Antofagasta, Avenida Angamos 601, Antofagasta 1270300, Chile\\
$^{3}$Leibniz-Institut f\"ur Astrophysik Potsdam, An der Sternwarte 16, D-14482 Potsdam, Germany\\
$^{4}$ICRAR, M468, The University of Western Australia, 35 Stirling Highway, Crawley, WA 6009, Australia\\
$^{5}$INAF-Osservatorio di Astrofisica e Scienza dello Spazio, Via Gobetti 93/3, I-40129 Bologna, Italy\\ 
$^{6}$School of Mathematical and Physical Sciences, Macquarie University, Balaclava Road, Sydney, NSW 2109, Australia\\
$^{7}$Research Centre for Astronomy, Astrophysics and Astrophotonics, Macquarie University, Balaclava Road, Sydney, NSW 2109, Australia\\
$^{8}$Istituto Nazionale di Fisica Nucleare (INFN)-Sez. di Napoli, Via Cinthia, 80126 Napoli, Italy\\ 
$^{9}$INAF–Osservatorio Astronomico di Padova, Vicolo dell’Osservatorio 5, I-35122 Padova, Italy\\
$^{10}$Koninklijke Sterrenwacht van Belgi\"e, Ringlaan 3, B-1180 Brussels, Belgium\\
$^{11}$European Southern Observatory, Ave. Alonso de Cordova 3107, Vitacura, Santiago 19001, Chile\\ 
$^{12}$European Southern Observatory, Karl-Schwarzschild-Str 2, D-85748 Garching bei Munchen, Germany\\ 
$^{13}$ Lund Observatory, Department of Astronomy and Theoretical Physics, Lund University, Box 43, 22100 Lund, Sweden\\
$^{14}$Lennard-Jones Laboratories, Keele University, ST5 5BG, UK
}

\date{Accepted XXX. Received YYY; in original form ZZZ}

\pubyear{2015}

\begin{document}
\label{firstpage}
\pagerange{\pageref{firstpage}--\pageref{lastpage}}
\maketitle

\begin{abstract}

We employed the {\it VISTA near-infrared $YJK_\mathrm{s}$ survey of the Magellanic System} (VMC), to analyse the $Y,\,J,\,K_\mathrm{s}$ light curves of $\delta$ Cepheid stars (DCEPs) in the Large Magellanic Cloud (LMC). Our sample consists of 4408 objects accounting for 97 per cent of the combined list of OGLE\,IV and {\it Gaia}\,DR2 DCEPs.
We determined a variety of period--luminosity ($PL$) and period--Wesenheit $PW$ relationships for Fundamental (F) and First Overtone (1O) pulsators. We discovered for the first time a break in these relationships for 1O DCEPs at $P$=0.58 d. 
We derived relative individual distances for DCEPs in the LMC with a precision of $\sim$1 kpc, calculating the position angle of the line of nodes and inclination of the galaxy: $\theta$=145.6$\pm$1.0 deg and $i$=25.7$\pm$0.4 deg. The bar and the disc are seen under different viewing angles.  
We calculated the ages of the pulsators, finding two main episodes
of DCEP formation lasting $\sim$40 Myr which happened 93 and 159 Myr ago. Likely as a result of its past interactions with the SMC, the LMC shows a non-planar distribution, with considerable structuring: the bar is divided into two distinct portions, the eastern and the western displaced by more than 1 kpc from each other. Similar behaviour is shown by the spiral arms. The LMC disc appears "flared" and thick, with a disc scale height of $h\sim 0.97$ kpc. This feature can be explained by strong tidal interactions with the Milky Way and/or the Small Magellanic Cloud or past merging events with now disrupted LMC satellites.

\end{abstract}




\section{Introduction}

The Large and the Small Magellanic Clouds (LMC and SMC) are the most massive satellite galaxies of the Milky Way (MW) and represent a fundamental benchmark in several  astrophysical fields. Indeed, due to their proximity \citep[$D_{\odot} \sim$50.0 and $\sim$62.5 kpc for the LMC and SMC, respectively:][]{Piet2019,Graczyk2020}, their stellar populations can be investigated down
to the intrinsically faintest and oldest constituents. 
Based on proper motions measurements \citep[][and references therein]{Kallivayalil2013} and the dynamical disturbance observed in the MW halo \citep{Conroy2021}, the Magellanic Clouds (MCs)  are thought to be on their first passage close to the MW. 
They are also important for the study of galaxy interactions. 
Evidences of their encounters are conspicuous: The Magellanic Stream is an elongated  neutral gas feature extending by over 200 deg on the sky, and is thought to be formed by gas stripped from the MCs during a close interaction about 2 Gyr ago \citep[see e.g.][and references therein]{Nidever2008,Nidever2010,Besla2012,Diaz2012,Hammer2015,Wang2019}.  
The Magellanic Bridge is a prominent structure constituted by gas and young stars connecting the SMC with the LMC \citep[see e.g.][and references therein]{Irwin1985,Demers1999,Noel2015,Piatti2015,Schmidt2020,Luri2021} and likely formed in a more recent interaction between the galaxies \citep[200--300 Myr ago, e.g.][and references therein]{Diaz2012,Besla2013,Wang2019}. Recent studies suggest that the stars in the Bridge are flowing from the SMC towards the LMC \citep[][]{Zivick2019,Schmidt2020,Luri2021}, further probing the nature of the interaction. 
Additional low surface brightness substructures resulting from interactions between the two MCs and between the MCs and the MW have been recently found in the outskirts of both the SMC and LMC by new observational campaigns \citep[for details see][and references therein]{Mackey2016,Mackey2018,Belokurov2019,Cullinane2021,Dalal2021,James2021}. Furthermore, the bodies of the SMC and LMC host the evidences of their mutual interaction. 
Indeed, the SMC shows a distorted shape and an extreme elongation along the line of sight, up to 20 kpc, as reported by many authors \citep[e.g.][and references therein]{Subramanian2015,Jacy2016,Ripepi2017}. Several studies also reported the presence of a stellar population tidally stripped from the inner SMC placed about 11 kpc closer to us \citep[][]{Subramanian2017,Dalal2021,James2021,Tatton2021}.
The LMC displays signatures of past interactions with the SMC and the MW in the three-dimensional (3D) geometry of the galaxy. It has been found that the bar of the LMC is offset from and slightly misaligned with the plane of the disc \citep[][]{Zhao2000,Nikolaev2004}. Additionally, the presence of a symmetric warp in the LMC disc and the bar being elevated above the disc plane were reported by  \citet[][]{Nikolaev2004}. Subsequent studies suggested that the disc is also truncated in the direction of the SMC \citep[e.g.][]{Choi2018,Mackey2018}.  

The study of the 3D geometry of the LMC is therefore crucial to understand the past interaction history of the galaxy. In addition, the study of the 3D geometry of the LMC allows us to estimate the LMC viewing angles, the inclination $i$ and the position angle of the lines of nodes $\theta$, i.e. the intersection of the galaxy and the sky planes. These parameters define the directions in which we observe the LMC disc. The measure of these angles has a significant impact on the determination of the dynamical state of the LMC, because they are used to transform the proper motions and line-of-sight velocities into rotational velocities, needed to obtain the stellar orbits \citep[see e.g.][]{Luri2021}.

The literature reports a variety of values for the LMC viewing angles. This is somehow expected because different populations sample different portions of the LMC, young and old populations have different geometrical distributions \citep[][]{Vaucouleurs1972,Cioni2000,vanderMarel2001,Weinberg2001,Moretti2014,Dalal2019}.

The viewing angles of the LMC have been studied using a variety of population tracers and methodologies. The old tracers adopted in the literature include red and asymptotic giant branch stars \citep[RGB and AGB;][]{Cioni2000}, red clump stars  \citep[RC;][]{Subramanian2013,Choi2018}, and RR Lyrae pulsating variables \citep[][]{Deb2014,Cusano2021}. Additional estimates have been obtained from the study of the dynamical properties of the above mentioned stellar tracers \citep[][]{vanderMarel2002,Olsen2011,Wan2020,Luri2021,Niederhofer2021} or from the spatially resolved star formation history (SFH) of the LMC \citep[][]{Mazzi2021}. The young stellar tracers are basically the $\delta$ Cepheid variables (DCEPs hereafter) for which precise distances can be derived by means of the period--luminosity and period--Wesenheit\footnote{The Wesenheit magnitudes are designed to be reddening-free by construction  \citep{Madore1982,Caputo2000} provided that the extinction law is known.} ($PL$ and $PW$ hereafter) relations that hold for these objects \citep[see e.g.][]{Leavitt1912,Madore1982,Caputo2000}. Thanks to their unique properties, the DCEPs have been extensively used to study the LMC disc in the literature \citep[e.g.][]{Nikolaev2004,Haschke2012,Jacy2016,Inno2016,Deb2018}. 
All these works provide viewing angle values that are not in agreement with one another by more than 10--15 per cent \citep[for a list of numerical values and related errors see e.g. Table 5, Table 3 and Table 2 of][respectively]{Inno2016,Mazzi2021,Niederhofer2021}, suggesting that further investigations are needed. 

In this work we take advantage of the time-series photometry in the $Y,J,\kk$ bands provided by the VISTA\footnote{Visible and Infrared Survey Telescope for Astronomy, https://www.eso.org/sci/facilities/paranal/telescopes/vista.html \citep[see also][]{Emerson2006}.} survey of the Magellanic Clouds \citep[VMC][]{Cioni2011} for the 97 per cent of known DCEPs present in the LMC to study the morphology of the young population and to estimate the viewing angles of the disc. 
With respect to previous works in the optical domain \citep[e.g.][]{Jacy2016} we have a better precision guaranteed by the lower intrinsic dispersion of the near-infrared (NIR hereafter) $PL$/$PW$ relations \citep[e.g.][]{Freedman2010, Caputo2000}. Furthermore, our photometry is more than one magnitude deeper than in previous works using NIR bands. This allowed us to measure even the faintest DCEPs in the LMC ($K_\mathrm{s,0} \sim 18$ mag), thus securing a higher completeness of the sample. It is also more homogeneous and precise, since we use well-sampled light-curves in the $\kk$ band from the VMC survey, while previous works generally relied on much less frequently sampled light curves.  

The paper is organized as follows: in Section 2 we describe the sample of DCEPs in the VMC survey; in Section 3 we analyse the light-curves and derive the intensity-averaged magnitudes in the $Y,J,\kk$ bands; in Section 4 we derive the $PL$/$PW$ relations for the DCEPs in the LMC; in Section 5 we calculate the ages of the DCEPs studied here; in Section 6 we investigate the 3D geometry of the LMC while in Section 7 we discuss the different substructures identified in the galaxy and analyse their dynamical properties in Section 8; Section 9 reports the conclusions.

\section{LMC Classical Cepheids in the VMC survey}

\begin{figure}
	\includegraphics[width=\columnwidth]{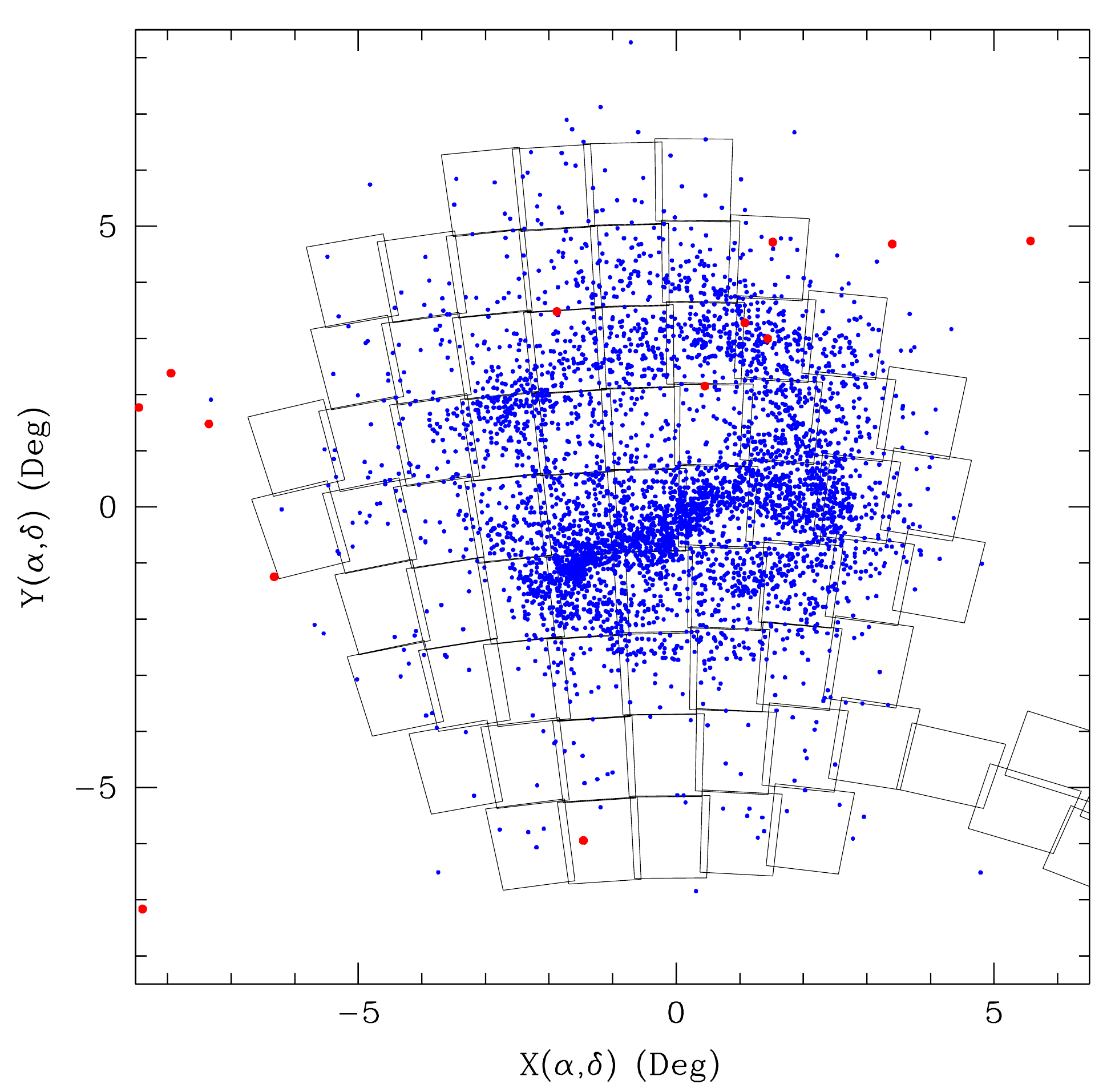}
    \caption{Map of the LMC DCEPs analysed in this paper. We used Cartesian coordinates obtained through a zenithal equidistant projection with centre $\alpha_0,\delta_0$=80.05,$-69.3$ deg (J2000). Blue and red points show the objects coming from the OGLE and {\it Gaia} catalogues, respectively. The red points were increased in size for a better visibility. Solid boxes represent the VISTA tiles building up the VMC survey.}
    \label{fig:map}
\end{figure}

The list of LMC DCEPs used as reference was 
taken from the OGLE\,IV survey \citep[Optical Gravitational Lensing Experiment\,IV;][]{Sos2017}, as updated in July 2019 on the OGLE website\footnote{http://www.astrouw.edu.pl/ogle/ogle4/OCVS/lmc/cep/}. Additional stars were taken from {\it Gaia} Data Release 1 \citep[DR1,][]{Prusti2016,Brown2016} and DR2 \citep[][]{Brown2018} lists \citep[see][for details]{Clementini2016,Clementini2019,Ripepi2019}. In more detail, OGLE\,IV published the identification, the $V,I$ light curves, and
main properties (periods, mean magnitudes, amplitudes etc.) for 4706 
DCEPs in the LMC. Similar data, but in the {\it Gaia} bands, was provided for 26 additional stars in {\it Gaia} DR1 and DR2. Most of these objects fall within the VMC survey footprint, as shown in Fig.~\ref{fig:map} where the area observed by the VMC survey in the region of the LMC is shown with contiguous squares representing one VMC tile each (the dimension of a tile is 1.77 deg$^2$ on the sky).
We cross-matched these samples with the VMC sources with 1 arcsec ~tolerance, retaining only objects having at least four epochs of observations in all the bands. In this way we ended up with 4495 and 5 useful light curves from the OGLE and {\it Gaia} samples, respectively. The overall completeness is $\sim$97 per cent. The classification of the DCEPs in terms of pulsation modes was taken from the OGLE and {\it Gaia} catalogues. Our sample counts 2390 F, 1705 1O and 405  
 mixed modes (F/1O, 1O/2O, F/1O/2O, 1O/2O/3O, 1O/3O), respectively.

\begin{figure}
	\includegraphics[width=\columnwidth]{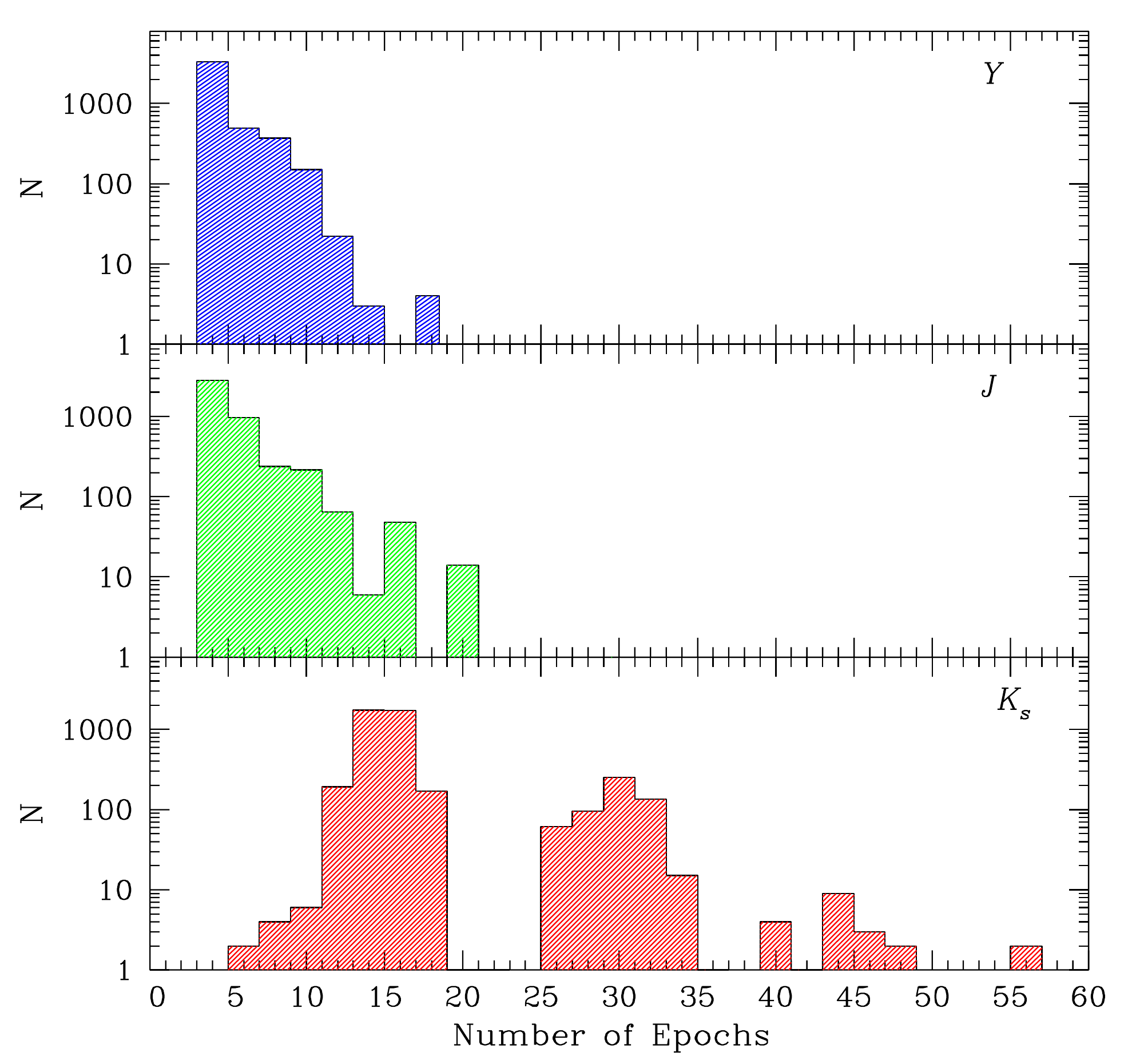}
    \caption{Histogram of the number of epochs in our LCs for each photometric band  (labelled in the figure).}
    \label{fig:nobs}
\end{figure}

A detailed description of the observations carried out in the context of the VMC
survey is reported in \citet{Cioni2011}. The procedures used to investigate the Cepheids of all types were discussed in detail in several papers of the collaboration \citep{Ripepi2012a,Ripepi2012b,Moretti2014,
Ripepi2014b,Ripepi2015,Moretti2016,Marconi2017,Ripepi2017}.
In brief, the VMC $K_\mathrm{s}$-band time--series observations were planned to obtain 
13 different epochs executed over several consecutive months with a min cadence of 1--3--5--7 and 17 days for 11 epochs in addition to two shallow  epochs (corresponding to half the exposure time of the other epochs) without specific cadence requirements. This 
observing strategy permitted to achieve well-sampled light curves for
 pulsating stars such as RR Lyrae and DCEPs.  
 In the case of the $Y$ and $J$ bands, we programmed to obtain four epochs, of which two are shallow. However, these were the minimum number of epochs expected in the light curves, as a few additional epochs became  
available for several tiles (particularly in the $K_\mathrm{s}$-band)
given that some Observing Blocks (OBs) were repeatedly executed to meet the required constraints on the sky conditions, but none the less provided usable data. 
In addition, the areas of overlap between the tiles include several DCEPs which therefore had twice the expected number of epochs. The number of epochs in each filter is shown in Fig.~\ref{fig:nobs}. On average we have 5.5$\pm$1.3 epochs in $Y$ and $J$, while in $K_\mathrm{s}$, we have a double-peaked distribution. The large majority of DCEPs ($\sim$ 3820 objects) have fewer than 20 epochs, with an average of 15.3$\pm$1.3, while $\sim$590 objects have more than 20 epochs, with an average of 30$\pm$4. 

The data used in this paper were processed by means of the pipeline version 1.5 of the
VISTA Data Flow System \citep[VDFS,][]{Emerson2004,Irwin2004}. The photometry 
is in the VISTA photometric system \citep[Vegamag=0; for details on the VISTA
photometric system see][]{Carlos2018}. The time--series data
used in this work were retrieved from the VISTA Science
Archive\footnote{http://horus.roe.ac.uk/vsa/} \citep[VSA,][]{Cross2012}. 

\begin{figure}
\vbox{
	\includegraphics[width=8.5cm]{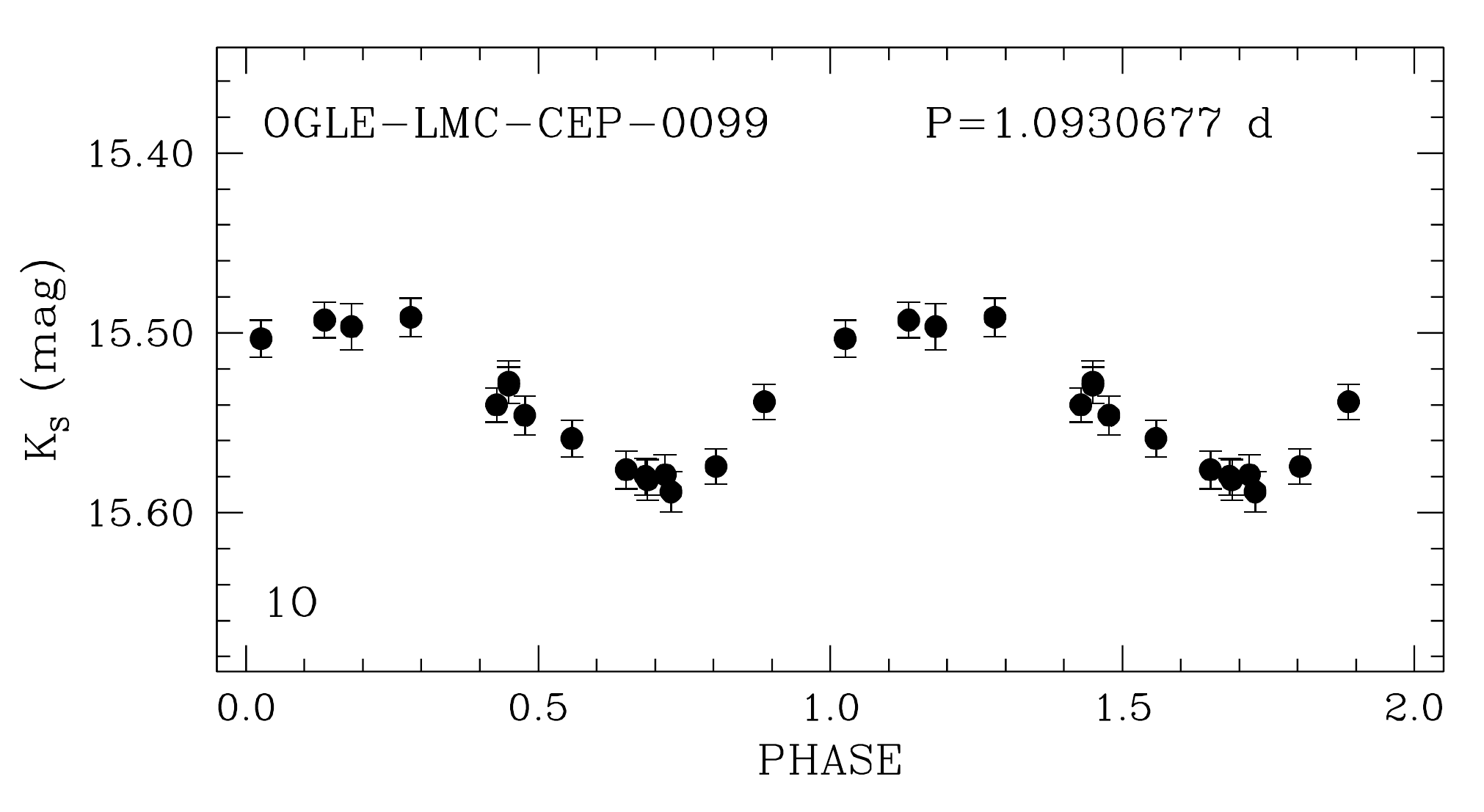}
	\includegraphics[width=8.5cm]{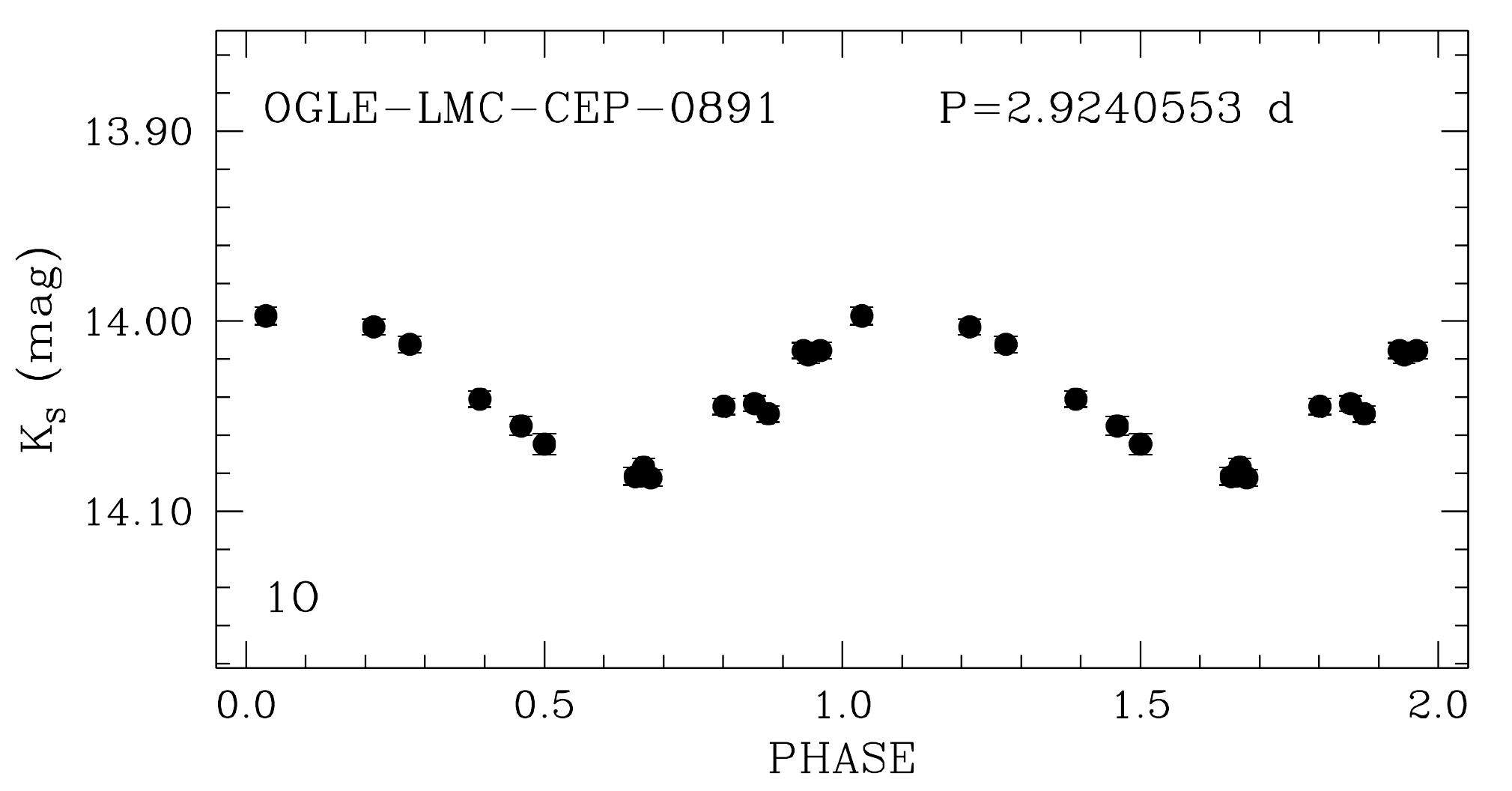}
		\includegraphics[width=8.5cm]{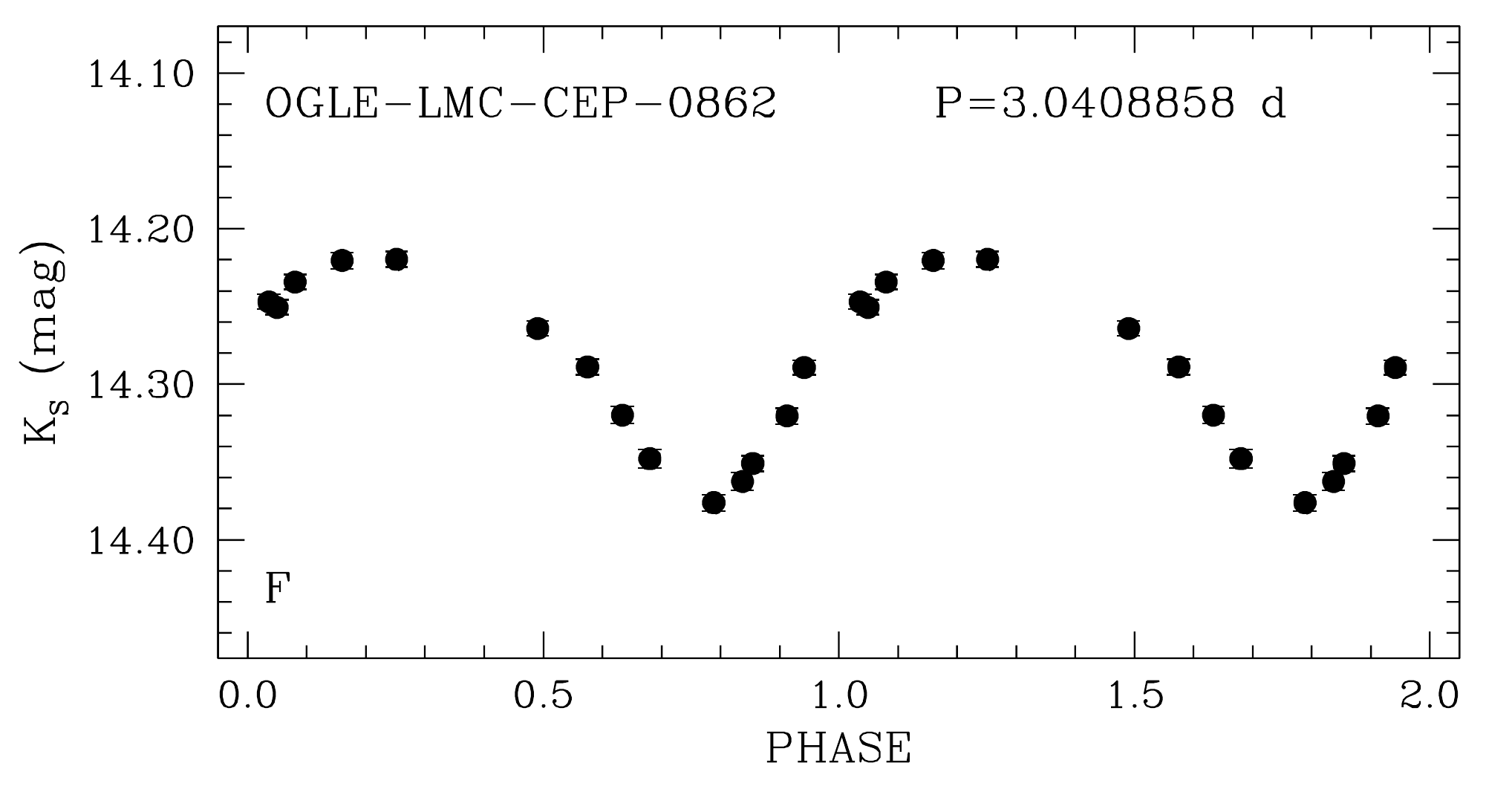}
			\includegraphics[width=8.5cm]{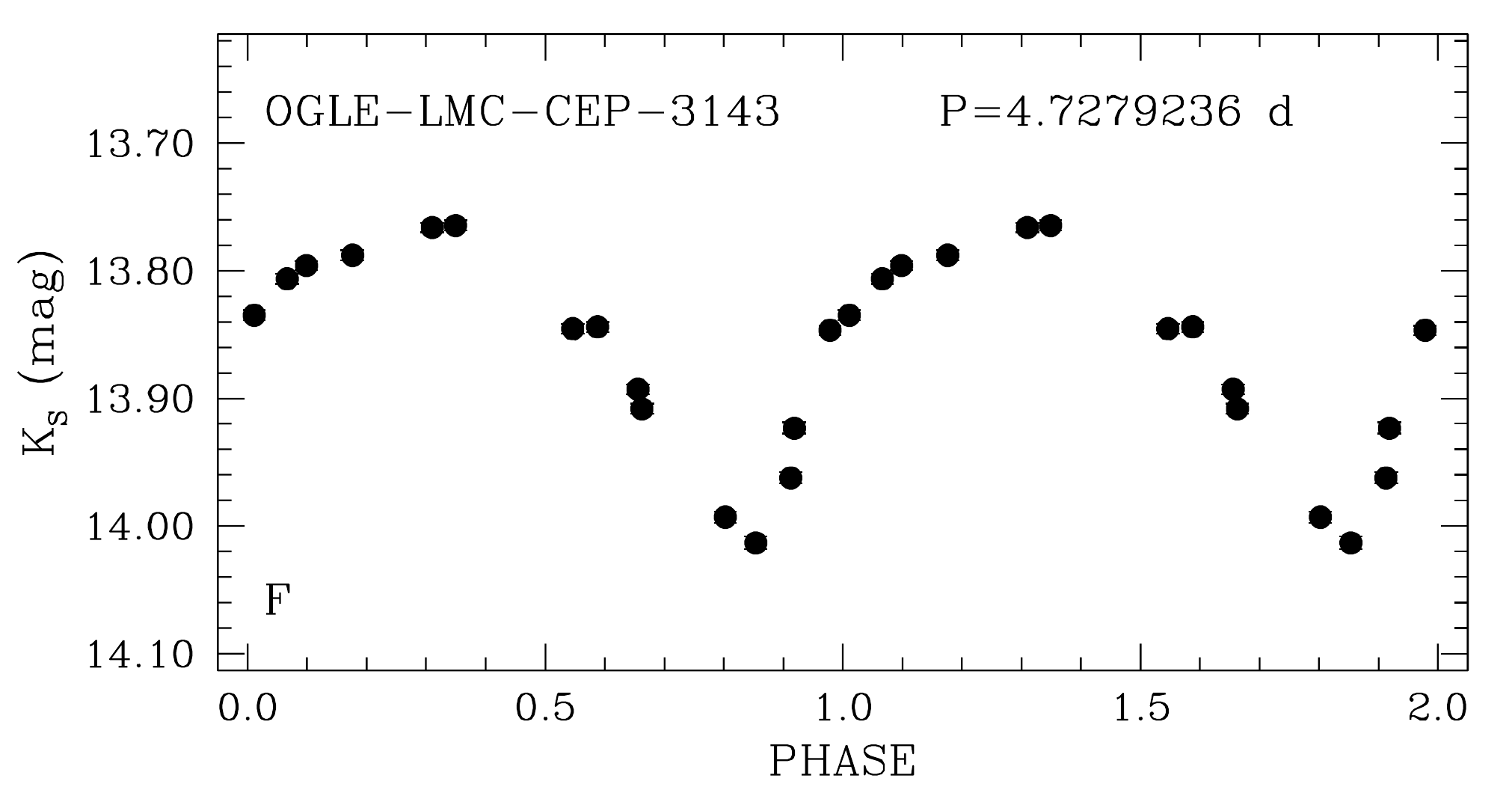}
	}
    \caption{Examples of light curves in the $\kk$ band. }
    \label{fig:LC}
\end{figure}

As in VMC the stars brighter than $K_\mathrm{s}\sim$12 mag can show saturation issues, we complemented our photometry with that by \citet{Persson2004} which includes 92 F-mode DCEPs with periods mainly in the range of 10--100 days. Since the $J$-band light curves (LCs) in \citet{Persson2004} are better sampled than ours, we used their photometry also for the few stars that have non-saturated VMC data. 
To homogenise \citet{Persson2004}'s photometry with ours, we first transformed their 
data from the Las Campanas Observatory (LCO) to the 2MASS system
using the relations by \citet{Carpenter2001} and then from 2MASS to the VISTA system using the relations given by \citet{Carlos2018}. 
Figure~\ref{fig:LC} shows the quality of our LCs, while Table~\ref{table:lcs} reports the $Y,J,\kk$ VMC time-series photometry used in this paper. The table with VMC photometry for the sample of 4408 DCEPs is provided electronically.

\begin{table}
\caption{$Y$, $J$ and $\kk$ time–series photometry for the 4408 LMC DCEPs 
investigated in this paper. The sample data below refer
to the variable OGLE-LMC-CEP-0020. The table is published in its entirety in the 
electronic version of the journal. \label{table:lcs}}
\begin{center}
\begin{tabular}{ccc}
\hline 
\noalign{\smallskip} 
HJD-2\,400\,000 & $Y$  & $\sigma_{Y}$  \\
     (d)       & (mag) & (mag) \\
\noalign{\smallskip}
\hline 
\noalign{\smallskip} 
   56687.57917  &  14.372  &   0.003 \\
   56688.59545  &  14.348  &   0.002\\
   56705.57569  &  14.393  &   0.002\\
   56712.57314  &  14.380  &   0.002\\
   56727.53181  &  14.390  &   0.002\\
\noalign{\smallskip}
\hline 
\noalign{\smallskip} 
HJD-2\,400\,000 & $J$  & $\sigma_{J}$  \\
     (d)       & (mag) & (mag) \\
\noalign{\smallskip}
\hline 
\noalign{\smallskip} 
   56695.61483  &  14.232  &   0.003\\
   56727.55304  &  14.223  &   0.002\\
   56967.69831  &  14.211  &   0.003\\
   56998.60337  &  14.198  &   0.003\\
\noalign{\smallskip}
\hline 
\noalign{\smallskip} 
HJD-2\,400\,000 & $K_\mathrm{s}$  & $\sigma_{K_\mathrm{s}}$  \\
     (d)       & (mag) & (mag) \\
\noalign{\smallskip}
\hline 
\noalign{\smallskip} 
   56673.54563  &  13.522  &   0.004\\
   56693.58784  &  13.854  &   0.004\\
   56697.53967  &  13.857  &   0.004\\
   56704.53190  &  13.875  &   0.004\\
   56712.54031  &  13.863  &   0.004\\
   56712.59488  &  13.865  &   0.004\\
   56728.52813  &  13.863  &   0.004\\
   56893.87320  &  13.884  &   0.004\\
   56967.72057  &  13.851  &   0.005\\
   56976.72364  &  13.867  &   0.004\\
   56998.62506  &  13.835  &   0.004\\
   57002.53873  &  13.974  &   0.005\\
   57034.59135  &  13.677  &   0.005\\
   57051.64906  &  13.868  &   0.004\\
   57069.57183  &  13.879  &   0.004\\
   57322.67660  &  13.851  &   0.004\\
   57709.68504  &  13.895  &   0.007\\
   57710.62212  &  13.918  &   0.005\\
   \noalign{\smallskip}
\hline 
\noalign{\smallskip}
\end{tabular}
\end{center}
\end{table}

\begin{landscape}
\begin{table*}
\begin{adjustwidth}{-7cm}{}
\footnotesize\setlength{\tabcolsep}{3pt} 
\caption{Photometric parameters for all the 4500 LMC DCEPs
  analysed in this paper. Columns: (1) Identification from OGLE\,IV; (2)  Mode: F=Fundamental; 1O=First
  Overtone; 2O=Second Overtone; 3O=Third Overtone; (3)--(4) RA and Dec; (5) $V$-band magnitude from OGLE; (6) Period; (7)--(8) Intensity-averaged magnitude in
  $Y$ and relative uncertainty; (9)--(10) Peak-to-peak amplitude in
  $Y$ and relative uncertainty; (11) to (14) As for columns (7) to (10)
  but for the $J$ band; (16) to (19) As for column (7) to (10) but for
  the $K_\mathrm{s}$ band; (19) $E(V-I)$ values adopted in this work; (20)-(25) flags indicating whether or not (0=included, 1=excluded) the star was used in the derivation of a specific $PL$/$PW$ relation. FL1 to FL3 refer to the $PL$ relations in the $Y_0,J_0$ and $K_\mathrm{s,0}$ magnitudes, respectively; FL4 to FL6 refer to $PW$ relations in $W(Y,K_\mathrm{s})$,$W(J,K_\mathrm{s})$ and  $W(V,K_\mathrm{s})$, respectively. This Table is published in its entirety in the electronic edition of the paper. A portion is
shown here for guidance regarding its form and content.}
\label{table:averages}
\begin{tabular}{crccccccccccccccccccccccc}
\hline  
\noalign{\smallskip}   
       OGLE\_ID    & MODE &  RA & Dec & $V$ &  $P$     & $\langle Y \rangle$ & $\sigma_{\langle Y \rangle}$&   A($Y$) &  $\sigma_{{\rm A}(Y)}$ & $\langle J \rangle$ & $\sigma_{\langle J \rangle}$&   A($J$) &  $\sigma_{{\rm A}(J)}$  & $\langle K_\mathrm{s} \rangle$ & $\sigma_{\langle K_\mathrm{s} \rangle}$&   A($K_\mathrm{s}$) &  $\sigma_{{\rm A}(K_\mathrm{s})}$ & $E(V-I)$ & FL1 & FL2 & FL3 & FL4 & FL5 & FL6  \\
   & & deg & deg & mag & days & mag &mag & mag & mag & mag & mag & mag & mag & mag & mag &   mag & mag & mag &  &  & &  & &  \\                     
(1)  & (2)  & (3) & (4) & (5) & (6) & (7) &(8) & (9) & (10) & (11) & (12)  & (13) & (14) & (15) &(16)  & (17) &(18) & (19) & (20) & (21)  & (22)  & (23)  & (24)  & (25)   \\                     
\noalign{\smallskip}
\hline  
\noalign{\smallskip}    
   OGLE-LMC-CEP-3367 &    1O/2O &  78.97004 & -69.76561  &  19.149 &  0.22042 &    18.202  &    0.0097  &     0.113  &     0.003  &    18.202  &    0.0097  &     0.113  &     0.003  &    18.002  &    0.0111  &     0.228  &     0.043 &  0.116 &  0 &  0 &   0  &  0 &   0  &  0  \\
   OGLE-LMC-CEP-3374 & 1O/2O/3O &  84.14108 & -68.10658  &  19.165 &  0.22226 &    18.262  &    0.0213  &     0.229  &     0.057  &    18.262  &    0.0213  &     0.229  &     0.057  &    18.063  &    0.0149  &     0.165  &     0.037 &  0.129 &  0 &  0 &   0  &  0 &   0  &  0  \\
   OGLE-LMC-CEP-4292 &    1O/2O &  85.51100 & -67.74036  &  18.710 &  0.23073 &    17.941  &    0.0022  &     0.110  &     0.004  &    17.941  &    0.0022  &     0.110  &     0.004  &    17.728  &    0.0110  &     0.126  &     0.020 &  0.112 &  0 &  0 &   0  &  0 &   0  &  0  \\
   OGLE-LMC-CEP-3369 & 1O/2O/3O &  80.29171 & -69.45181  &  18.864 &  0.23187 &    18.291  &    0.0486  &     0.066  &     0.095  &    18.291  &    0.0486  &     0.066  &     0.095  &    18.196  &    0.0327  &     0.259  &     0.124 &  0.053 &  0 &  0 &   0  &  1 &   0  &  1  \\
   OGLE-LMC-CEP-4706 &    1O/2O &  90.86062 & -67.10144  &  19.002 &  0.23373 &    18.214  &    0.0084  &     0.199  &     0.041  &    18.214  &    0.0084  &     0.199  &     0.041  &    17.932  &    0.0181  &     0.115  &     0.047 &  0.058 &  0 &  0 &   0  &  0 &   0  &  0  \\
   OGLE-LMC-CEP-1708 &    1O/2O &  80.45175 & -69.83333  &  18.693 &  0.24199 &    17.999  &    0.0353  &     0.175  &     0.057  &    17.999  &    0.0353  &     0.175  &     0.057  &    17.748  &    0.0268  &     0.197  &     0.056 &  0.131 &  0 &  0 &   0  &  0 &   0  &  0  \\
   OGLE-LMC-CEP-3878 & 1O/2O/3O &  80.02175 & -66.67419  &  18.748 &  0.24435 &    17.964  &    0.0014  &     0.206  &     0.005  &    17.964  &    0.0014  &     0.206  &     0.005  &    17.744  &    0.0117  &     0.083  &     0.037 &  0.133 &  0 &  0 &   0  &  0 &   0  &  0  \\
   OGLE-LMC-CEP-3440 &    1O/2O &  73.27383 & -66.20814  &  19.165 &  0.24460 &    18.266  &    0.0102  &     0.112  &     0.014  &    18.266  &    0.0102  &     0.112  &     0.014  &    18.010  &    0.0207  &     0.085  &     0.035 &  0.075 &  0 &  0 &   0  &  0 &   0  &  0  \\
   OGLE-LMC-CEP-4707 &    1O/2O &  91.49279 & -66.51028  &  18.720 &  0.24678 &    17.912  &    0.0089  &     0.155  &     0.023  &    17.912  &    0.0089  &     0.155  &     0.023  &    17.708  &    0.0131  &     0.101  &     0.060 &  0.057 &  0 &  0 &   0  &  0 &   0  &  0  \\
   OGLE-LMC-CEP-3563 &       1O &  76.22429 & -64.46194  &  18.579 &  0.25372 &    17.826  &    0.0051  &     0.127  &     0.014  &    17.826  &    0.0051  &     0.127  &     0.014  &    17.630  &    0.0082  &     0.103  &     0.028 &  0.045 &  0 &  0 &   0  &  0 &   0  &  0  \\
\noalign{\smallskip}
\hline  
\noalign{\smallskip}
\end{tabular}

\end{adjustwidth}
\end{table*}
\end{landscape}

\section{Data analysis}

We measured the $Y,~J$ and $\kk$ intensity-averaged magnitudes and the peak-to-peak amplitudes for the 4408 DCEPs considered in this work using a technique similar to that of our previous papers \citep[][]{Ripepi2016,Ripepi2017}. 
The basic idea is to construct a series of templates for each
of the three $Y,~J$ and $\kk$ bands and to use a modified
$\chi^2$ technique to determine the best-fitting template. 
The detailed procedure is outlined in the following sections. 

\subsection{Template derivation}

\begin{figure}
	\includegraphics[width=\columnwidth]{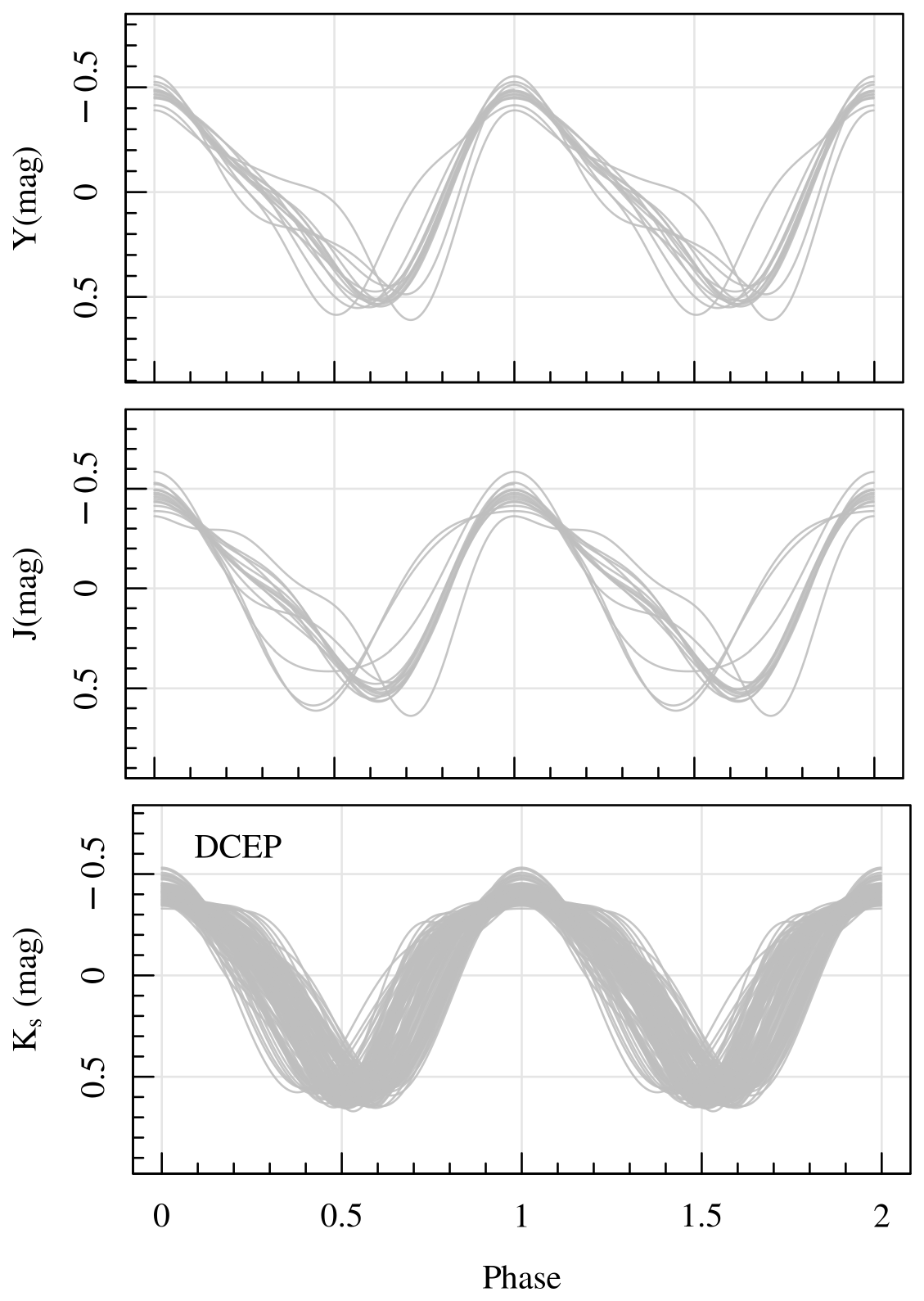}
    \caption{From top to bottom grey lines show the adopted light curve templates for the LMC DCEPs in the $Y$, $J$ and $K_\mathrm{s}$ bands, respectively.}
    \label{fig:templates}
\end{figure}

To construct the template curves, we have modelled all photometric time series in the $Y$, $J$, $\kk$ by using a proprietary C--code which fits light curves with a Fourier series automatically truncated at a given number of harmonics $N$, fixed by a statistical F-test:
\begin{equation}
    m(\phi) = A_0 + \sum_1^N A_i \cos(2\pi \phi + \Phi_i).
\end{equation}
Then we visually inspected all  light-curves fitted with more than 2 harmonics and selected a sample of the best fits trying to include the largest variety possible for what concerns periods and shapes of the light-curves. The selected models have been transformed into template models by performing the following steps: i) each model was subtracted from its intensity-averaged mean magnitude; ii) each model was re-scaled for its peak-to-peak amplitude. After this procedure, all templates consisted of light-curves with zero mean value and peak-to-peak amplitude equal to 1. The final template sample contains 14, 15 and 133 models respectively for $Y$, $J$ and $\kk$ bands. They are collectively shown in Fig.~\ref{fig:templates}.

\subsection{Fit to the data with templates}

The templates described above have been used to estimate the mean magnitudes and amplitudes of the selected sample of sources. Modelling data with templates instead of e.g. truncated Fourier series, allows us to avoid model oscillations, in particular for those light curves containing few observations. 

First, we folded the data according to the period and epoch values from the literature. We adopted the ephemerides from \citet{Sos2017} and \citet{Clementini2019,Ripepi2019} for the DCEPs identified by OGLE IV and {\it Gaia}, respectively.  
Then, for every light curve in a given filter we searched for the best template describing the observations. Each template was fitted to the observations by shifting it in magnitude and phase and scaling it to match the observed light curve amplitude. From the computational point of view this is obtained by minimising the following $\chi^2$ function:
\begin{equation}\label{chi}
\chi^2=\sum_{i}^{N_{\textrm{pts}}}\frac{[m_i-(a\cdot\textit{M}_{{t}}(\phi_i+\delta \phi)+\delta M)]^2}{\sigma_i^2}
\end{equation}
where the sum is over the number of epochs, $N_{\rm pts}$, the observed magnitudes are indicated with $m_i$ and their corresponding phases with $\phi_i$, while $M_t(\phi)$ is a smoothing spline function modelling the template curve. In Eq.~\ref{chi} the fitted parameters are the magnitude shift, $\delta M$, the scaling factor, $a$, and the phase shift $\delta \phi$.

The fitting routine also allows to reject outliers. They are identified by looking at the distribution of the residuals from the fit and by flagging points which are outside the  interval $-3.5\cdot DMAD$ to $+3.5\cdot DMAD$, where $DMAD$ stands for double-MAD, i.e. the Median Absolute Deviation calculated by treating separately the values smaller and larger than the median of the considered distribution. We limited also the maximum fraction of rejected points to 30 per cent of the time series length. 

Best fitting templates were then selected using the goodness parameter $G$ introduced by \citet{Ripepi2016}, for which here we adopted a slightly different definition: 
\begin{equation}
G = \left ( \frac{1}{\sigma}\right )^2 \cdot \left (\frac{N_U}{N_T}\right )^4    
\end{equation}
where $\sigma$ is the rms of the residuals, $N_U$ is the effective number of points used in the fit (i.e. after the rejection of outliers) and $N_T$ is the initial (i.e. including outliers) number of points. 
The best fitting template is the one corresponding to the maximum value of the $G$ parameter. 
Note that we used all the templates for F, 1O and mixed mode pulsators, and let the pipeline decide the best model, irrespective of the template provenience. We verified by means of visual inspection of the data that this strategy allows us to obtain better results compared with the imposition of templates with periods close to those of the analysed light curves. As a quantitative verification of this methodology, we calculated for each star the dispersion in magnitude obtained resulting from the fit with all the templates in a particular band. As a result of this exercise, we find that in the $\kk$ band 98\% of the stars have a dispersion lower than 1\%. This is not unexpected, as in the $\kk$ band we have more epochs of observations than in $Y$ or $J$ bands. As for the latter, the lower number of epochs makes the results a bit worse, but in any case for 74\% and 97\% of the stars we have a dispersion $<2$\% and $<5$\%, respectively. More importantly, there is no systematic trend with the period or pulsation mode, including for mixed mode pulsators.

To estimate the uncertainties on the fit parameters we used a Monte Carlo approach similar to \citet{Ripepi2016}. Briefly, for each fitted source we generated 100 bootstrap simulations of the observed time series. Then the template fitting procedure was repeated for each mock time series and a statistical analysis of the obtained fitted parameters was performed. Our fitted parameters error estimate is given by the robust standard deviation ($1.4826\cdot MAD$) of the distributions obtained by the quoted bootstrap simulations.

\begin{figure}
	\includegraphics[width=\columnwidth]{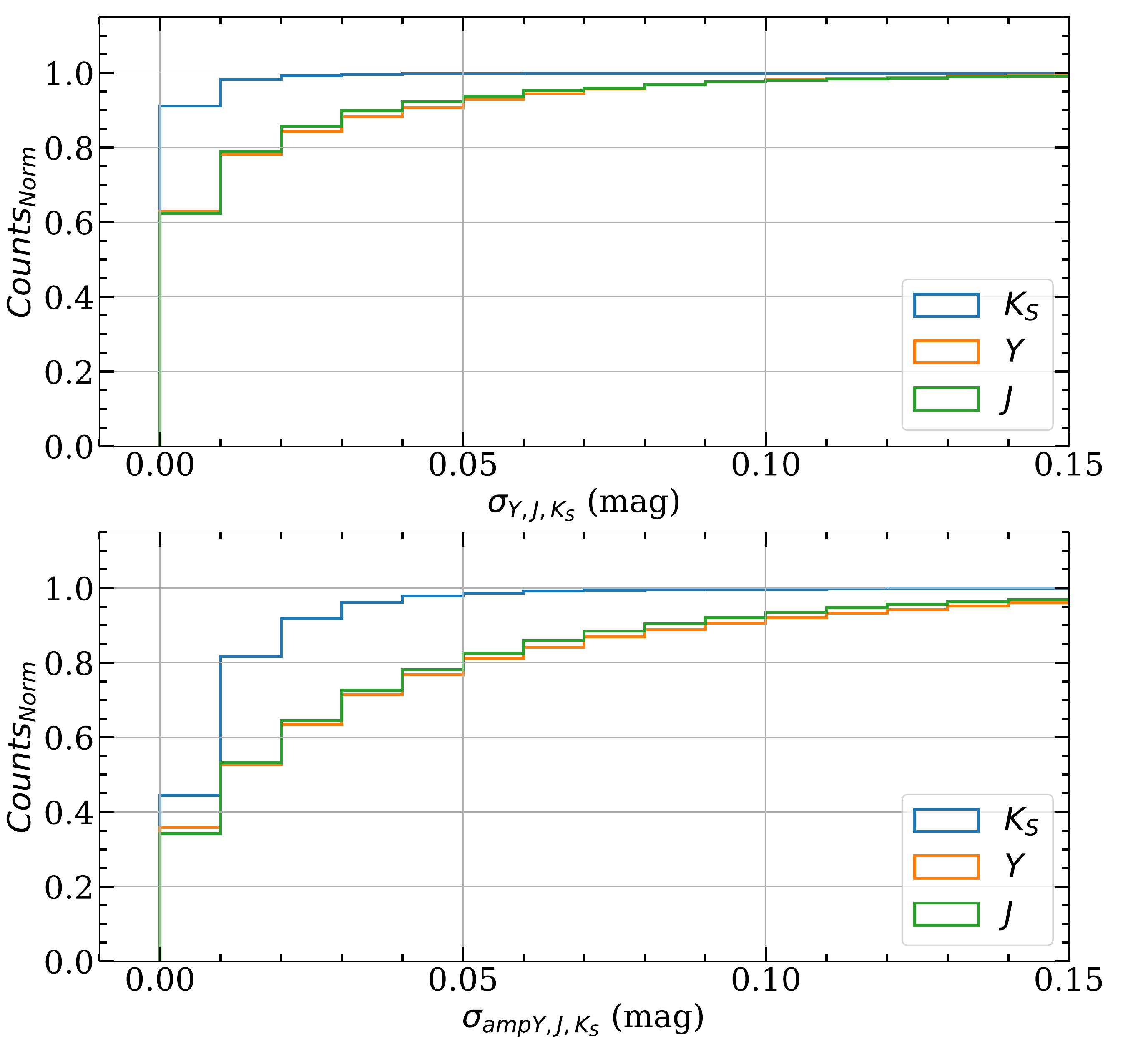}
    \caption{Top panel: distribution of the errors on $Y$,$J$,$K_\mathrm{s}$ magnitudes according to the bootstrap technique adopted in this work; Bottom panel: as in the top panel but for amplitudes.}
    \label{fig:errors}
\end{figure}

\begin{figure}
	\includegraphics[width=\columnwidth]{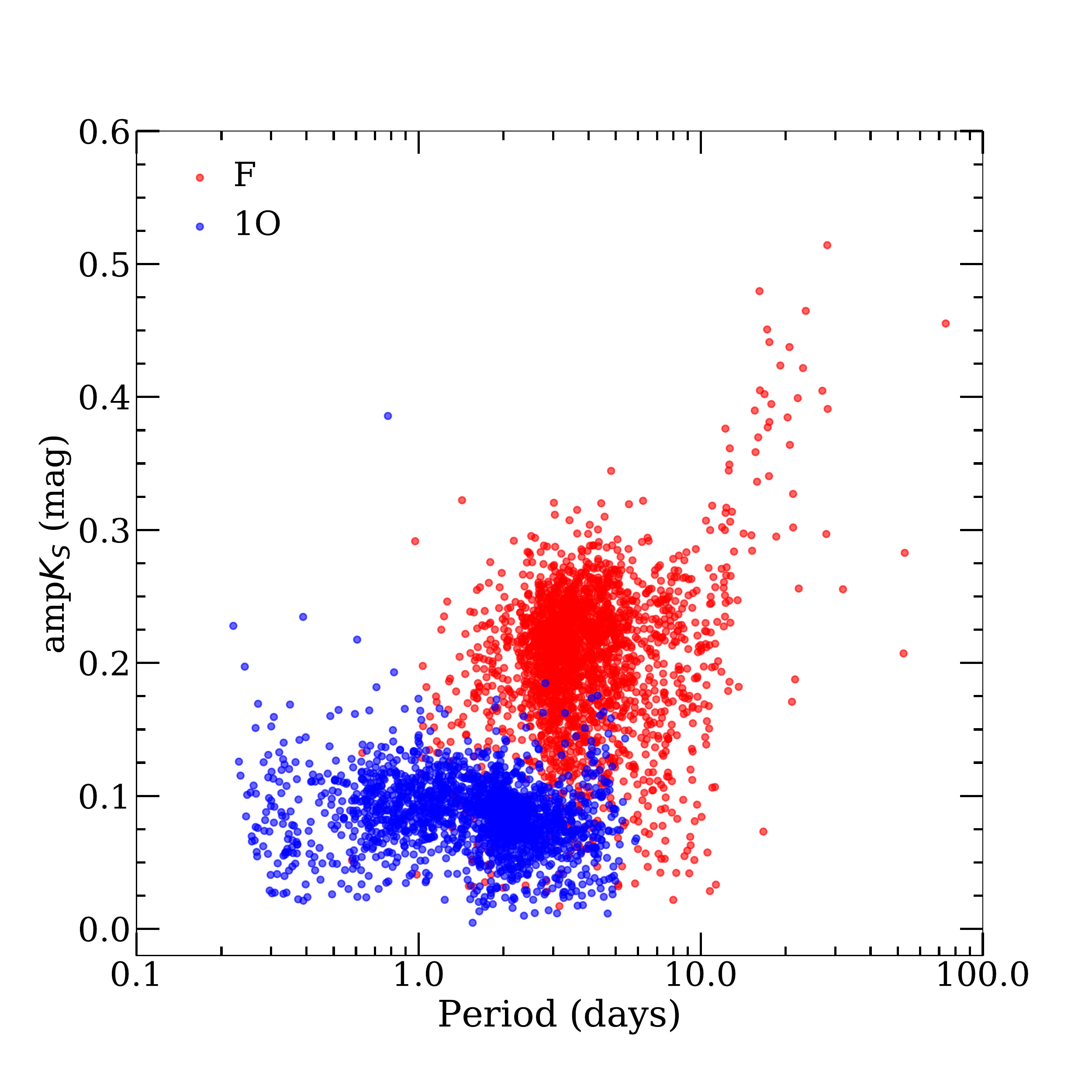}
    \caption{Period--amplitude in $K_\mathrm{s}$ band distribution for the LMC DCEPs investigated here.}
    \label{fig:amplitudes}
\end{figure}

The error distributions of the intensity-averaged magnitudes and peak-to-peak amplitudes are shown in Fig.~\ref{fig:errors}. In the $\kk$ band $\sim$90 per cent and $\sim$99 per cent of the DCEPs have errors on the intensity-averaged magnitudes lower than 0.01 and 0.02 mag, respectively. These percentages become $\sim$80 per cent and $\sim$87 per cent for the $Y$ and $J$ bands, owing to the significantly smaller number of epochs available in these bands. The bottom panel of Fig.~\ref{fig:errors} shows the same data but for the  peak-to-peak amplitudes. In this case the errors are, as expected, slightly larger than for the intensity-averaged magnitudes, but smaller than a few  per cent for the large majority of the stars. The goodness of the $\kk$ band peak-to-peak amplitudes can be appreciated in Fig.~\ref{fig:amplitudes}, where the loci of F and 1O DCEPs are shown, separated as expected \citep[compare with the SMC,][]{Ripepi2016}.

The $Y$, $J$ and $K_\mathrm{s}$ intensity-averaged magnitudes and amplitudes derived with the above procedure for the 4408 LMC DCEPs analysed in this paper are provided in Table~\ref{table:averages}, along with the respective errors. For completeness, Table~\ref{table:averages} also reports the data for the 92 sources whose photometry was taken from \citet{Persson2004}.

\begin{table}
\scriptsize
\caption{$PL$ and $PW$ relations for DCEP\_F and DCEP\_1O. The
Wesenheit functions are defined in the table. The columns "n" and "out" show the number of used and rejected objects, respectively.\label{table:pl}}
\begin{center}
\begin{tabular}{lccccccc}
\hline 
\noalign{\smallskip} 
Mode    &   $a$ & $\sigma a$&   $b$ & $\sigma b$ & rms & $n$ & out\\
\noalign{\smallskip} 
\noalign{\smallskip} 
\hline 
\noalign{\smallskip} 
\multicolumn{8}{c}{$Y_0$=$a$ log$P$+$b$}\\
\noalign{\smallskip} 
\hline 
\noalign{\smallskip}  
F                 &  $-$3.023  &  0.012 &   13.436  &   0.006  &   0.145   &   2416 &   38 \\
1O $P<0.58$ d     &  $-$3.786  &  0.112 &   17.313  &   0.013  &   0.148   &    135 &   23 \\
1O $P\ge0.58$ d   &  $-$3.291  &  0.015 &   14.982  &   0.003  &   0.141   &   1815 &   29 \\
\noalign{\smallskip} 
\hline 
\noalign{\smallskip} 
\multicolumn{8}{c}{$J_0$=$a$ log$P$+$b$}	      \\		                         
\noalign{\smallskip} 
\hline 
\noalign{\smallskip} 
F                 &  $-$3.084  &  0.010 &   13.199  &   0.005  &   0.127   &   2436 &   47 \\ 
1O $P<0.58$ d     &  $-$3.773  &  0.113 &   17.162  &   0.013  &   0.151   &    142 &   16 \\
1O $P\ge0.58$ d   &  $-$3.319  &  0.013 &   14.800  &   0.003  &   0.126   &   1829 &   30 \\
\noalign{\smallskip} 
\hline 
\noalign{\smallskip} 
\multicolumn{8}{c}{$K_\mathrm{s,0}$=$a$ log$P$+$b$}\\
\noalign{\smallskip} 
\hline 
\noalign{\smallskip} 
F                 &  $-$3.230  &  0.007 &   12.785  &   0.003  &   0.088   &   2427 &   56 \\
1O $P<0.58$ d     &  $-$3.867  &  0.107 &   16.939  &   0.012  &   0.143   &    138 &   20 \\
1O $P\ge0.58$ d   &  $-$3.402  &  0.011 &   14.504  &   0.002  &   0.099   &   1818 &   41 \\
\noalign{\smallskip} 
\hline 
\noalign{\smallskip} 
\multicolumn{8}{c}{$W(Y,K_\mathrm{s})$=$K_\mathrm{s}-0.42\,(Y-K_\mathrm{s})$=$a$ log$P$+$b$}\\
\noalign{\smallskip} 
\hline 
\noalign{\smallskip} 
F                 &  $-$3.306  &  0.007 &   12.516  &   0.003  &   0.082   &   2395 &   66 \\
1O $P<0.58$ d     &  $-$3.862  &  0.085 &   16.805  &   0.010  &   0.111   &    127 &   33 \\
1O $P\ge0.58$ d   &  $-$3.454  &  0.010 &   14.304  &   0.002  &   0.093   &   1797 &   54 \\
\noalign{\smallskip} 
\hline 
\noalign{\smallskip} 
\multicolumn{8}{c}{$W(J,K_\mathrm{s})$=$K_\mathrm{s}-0.69\,(J-K_\mathrm{s})$=$a$ log$P$+$b$}\\
\noalign{\smallskip} 
\hline 
\noalign{\smallskip} 
F                 &  $-$3.314  &  0.007 &   12.505  &   0.003  &   0.088   &   2424 &   66 \\
1O $P<0.58$ d     &  $-$3.943  &  0.107 &   16.796  &   0.013  &   0.143   &    135 &   26 \\
1O $P\ge0.58$ d   &  $-$3.457  &  0.010 &   14.300  &   0.002  &   0.093   &   1804 &   63 \\
\noalign{\smallskip} 
\hline 
\noalign{\smallskip} 
\multicolumn{8}{c}{$W(V,K_\mathrm{s})$=$K_\mathrm{s}-0.13\,(V-K_\mathrm{s})$=$a$ log$P$+$b$}\\
\noalign{\smallskip} 
\hline 
\noalign{\smallskip} 
F                 &  $-$3.303  &  0.007 &   12.564  &   0.003  &   0.079   &   2306 &   58 \\
1O $P<0.58$ d     &  $-$3.866  &  0.088 &   16.831  &   0.010  &   0.114   &    128 &   27 \\
1O $P\ge0.58$ d   &  $-$3.442  &  0.010 &   14.345  &   0.002  &   0.089   &   1714 &   52 \\
\noalign{\smallskip}
\hline
\end{tabular}
\end{center}
\end{table}

\begin{figure}
	\includegraphics[width=\columnwidth]{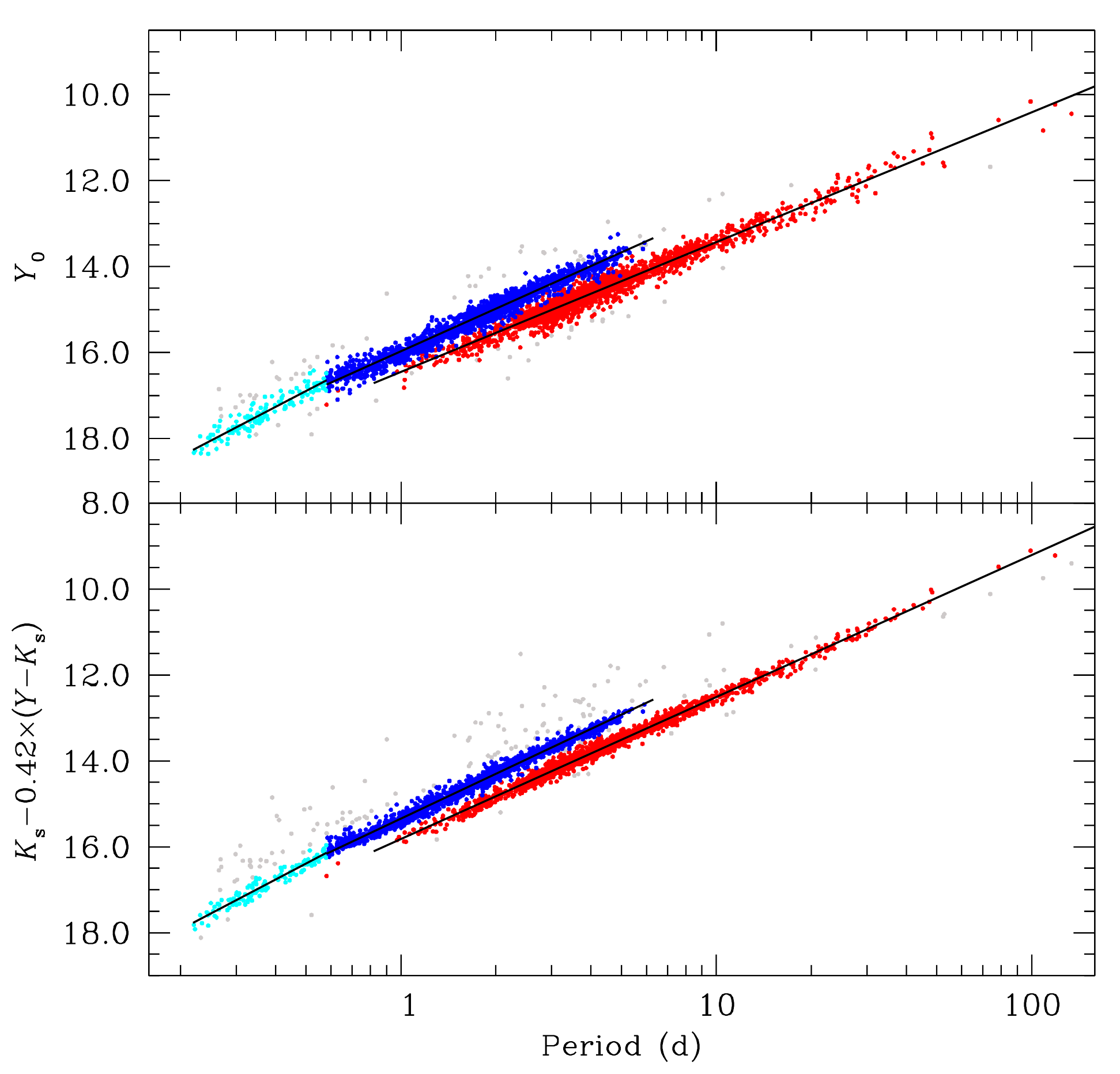}
    \caption{$PLY$ and $PWY\kk$ relations for our LMC DCEP sample. Red, blue and light blue dots show F-mode and 1O-mode with periods longer or shorter than 0.58 d, respectively. Grey dots are the outliers not included in the fit. }
    \label{fig:figY}
\end{figure}

\begin{figure}
	\includegraphics[width=\columnwidth]{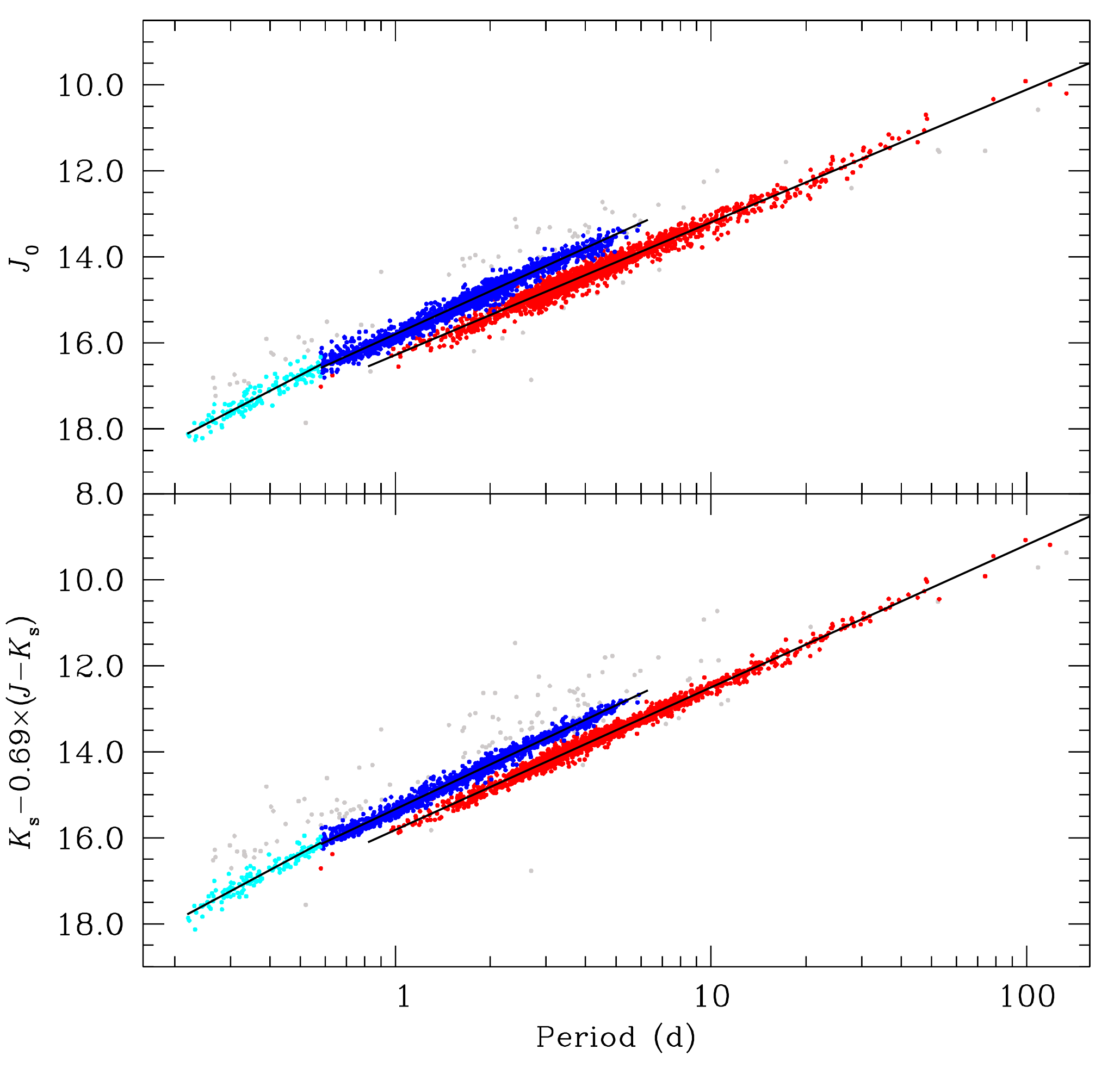}
    \caption{As in Fig.~\ref{fig:figY} but for the $PLJ$ and $PWJ\kk$ relations. }
    \label{fig:figJ}
\end{figure}

\begin{figure}
	\includegraphics[width=\columnwidth]{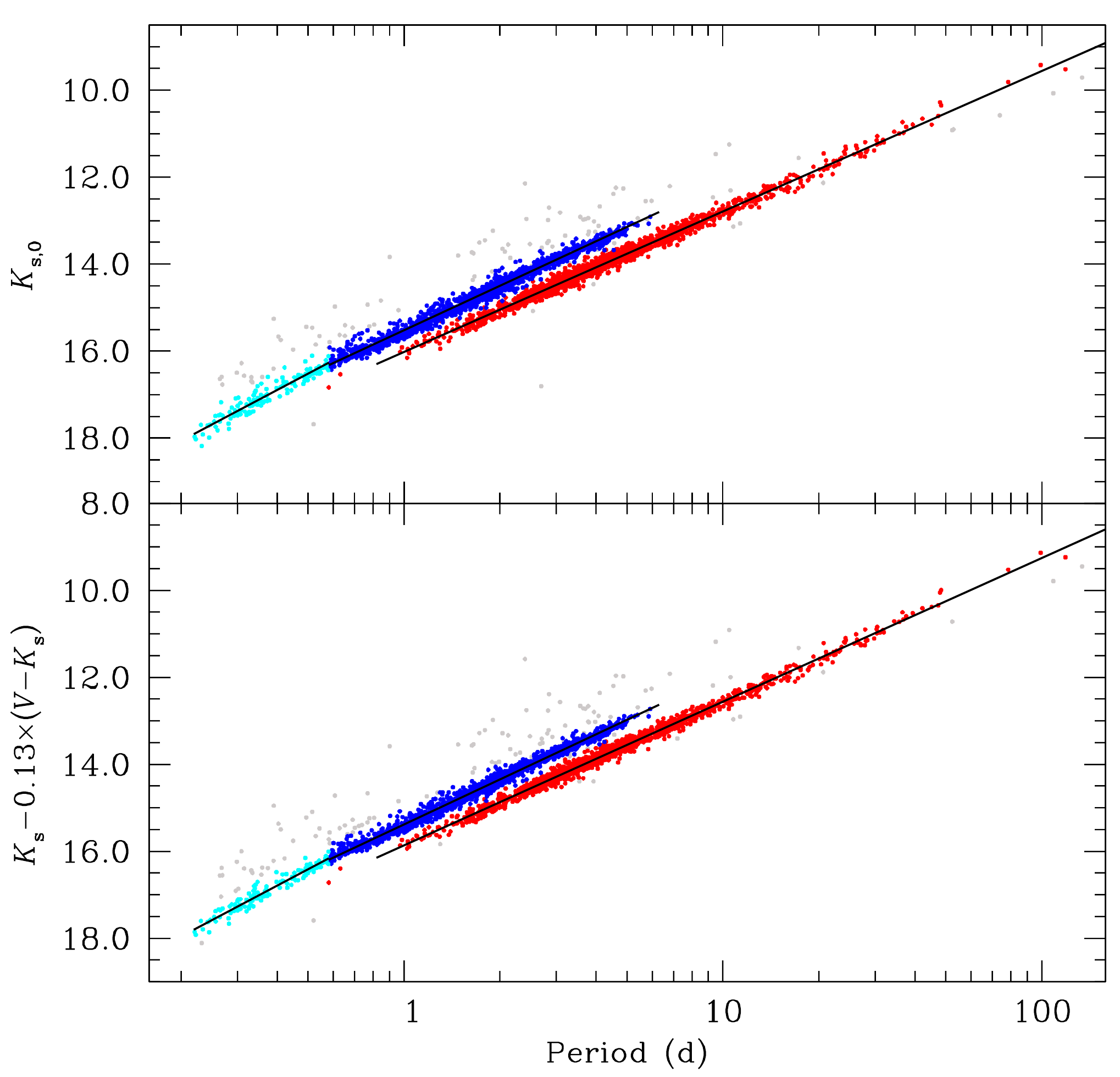}
    \caption{As in Fig.~\ref{fig:figY} but for the $PL\kk$ and $PWV\kk$ relations. }
    \label{fig:figK}
\end{figure}

\begin{table}
\caption{Inclination and position angle of the line of nodes for DCEP\_F, DCEP\_1O and the joint DCEP\_F and DCEP\_1O~samples (shown in the table with F+1O) obtained with the labelled $PL$ or $PW$ relations. The number of pulsators used in each calculation is also shown.
\label{table:inclination}}
\begin{center}
\begin{tabular}{lcccc}
\hline 
\noalign{\smallskip} 
$PL$ or $PW$ &  mode & Inclination    &  Pos. Angle & $n$ \\
             &        & (deg)    &  (deg)   &  \\
\noalign{\smallskip} 
\hline 
\noalign{\smallskip} 
$PLJ$          & F     &    26.4$\pm$1.4  &  162.2$\pm$3.5    &     2439    \\ 
$PLJ$          & 1O    &    25.2$\pm$1.7  &  145.4$\pm$4.2    &     1830    \\
$PLJ$          & F+1O  &    25.7$\pm$1.1  &  154.8$\pm$2.7    &     4269    \\
$PL\kk$        & F     &    25.9$\pm$0.9  &  150.6$\pm$2.2    &     2433    \\
$PL\kk$        & 1O    &    25.7$\pm$1.1  &  140.9$\pm$2.7    &     1826    \\
$PL\kk$        & F+1O  &    25.3$\pm$0.7  &  146.2$\pm$1.8    &     4259    \\
$PWJ\kk$       & F     &    25.4$\pm$1.0  &  148.0$\pm$2.5    &    2424     \\
$PWJ\kk$       & 1O    &    26.6$\pm$1.0  &  141.0$\pm$2.6    &    1804     \\
$PWJ\kk$       & F+1O  &    25.9$\pm$0.7  &  144.6$\pm$1.8    &    4228     \\
$PWV\kk$       & F     &    26.0$\pm$0.9  &  150.1$\pm$2.2    &    2269     \\
$PWV\kk$       & 1O    &    26.1$\pm$1.0  &  140.5$\pm$2.6    &    1714     \\
$PWV\kk$       & F+1O  &    25.8$\pm$0.7  &  146.0$\pm$1.7    &    3983     \\
\noalign{\smallskip}
\hline
\end{tabular}
\end{center}
\end{table}

\section{Period--Luminosity and Period--Wesenheit relations}

The intensity-averaged magnitudes estimated as described in the previous section were used in conjunction with the literature periods to construct new $PL$ and $PW$ relations for the LMC~DCEPs that are at the basis of our structural analysis of the galaxy. The mixed mode DCEPs were used adopting as period the longest one, so that for example, F/1O and 1O/2O pulsators contribute to the determination of $PL$ or $PW$ relations in the F and 1O modes, respectively.
To calculate the $PL$ relations we first calculated the dereddened $Y,~J$ and
$K_\mathrm{s}$ magnitudes. To this aim we adopted the recent high-resolution (1.7 arcmin $\times$ 1.7 arcmin) empirically-calibrated reddening maps by \citet{Skowron2021} providing $E(V-I)$ values based on the RC stars. We choose these maps for their solid empirical calibration, homogeneity and ease of use, as well as for the good agreement with a variety of previous determinations, obtained with a broad range of different methods. For the extinction corrections, as in our previous works \citep[e.g.][]{Ripepi2016}, we used the coefficients from  \citet{Cardelli1989,Kerber2009,Gao2013}. In total, we investigated three $PL$ relations, namely $PLJ$, $PLY$, and $PLK_\mathrm{s}$. Similarly, we calculated the following $PW$ relations: $PWJ\kk=\kk-0.69 \times (J-\kk)$, $PWV\kk=\kk-0.13 \times (V-\kk)$ and $PWY\kk=\kk-0.42 \times (Y-\kk)$, where the coefficients of the three relations have been taken from the papers quoted above.   

To fit the relations we adopted the traditional least-squares (LS) fitting procedure, using a conservative sigma-clipping algorithm (3.5$\sigma$) to ensure removing only stars with actual problems with the data. In fact, several stars (about 3 per cent) are found above the expected position on the $PL$ and $PW$ relations, an occurrence in most cases due to blending, caused by the high crowding of certain regions of the LMC. An alternative  explanation could be the presence of several binary DCEPs with red giant companions \citep[][]{Pilecki2021}.
In performing the LS procedure, we soon realised that the 1O pulsators could not be fitted with just a single linear fit across the entire period range. Indeed, we found a break in all the $PL$ and $PW$ relations at $P$=0.58$\pm$0.1 d. In particular, the break was well visible in the $PL\kk$ and in the $PWV\kk$ relations, which rely on solid data for what concerns both the $\kk$ (see Sect. 3.2) and the OGLE\,IV $V$ magnitudes, and therefore it is definitely a real feature. The precise period location of the break was found experimenting with different thresholds and retaining the value of breaking period that provided the lowest root mean square (rms) for the $PL$ and $PW$ relations on both sides of the break. 
We will return later to the origin of such a break. Instead, no break was found for the F pulsators. 
The new PL, PW relations we have derived are shown in Figs.~\ref{fig:figY},~\ref{fig:figJ} and \ref{fig:figK}, their coefficients are summarised in Table~\ref{table:pl}. 

\subsection{On the origin of the break in the $PL/PW$ relations}

\begin{figure}
	\includegraphics[width=\columnwidth]{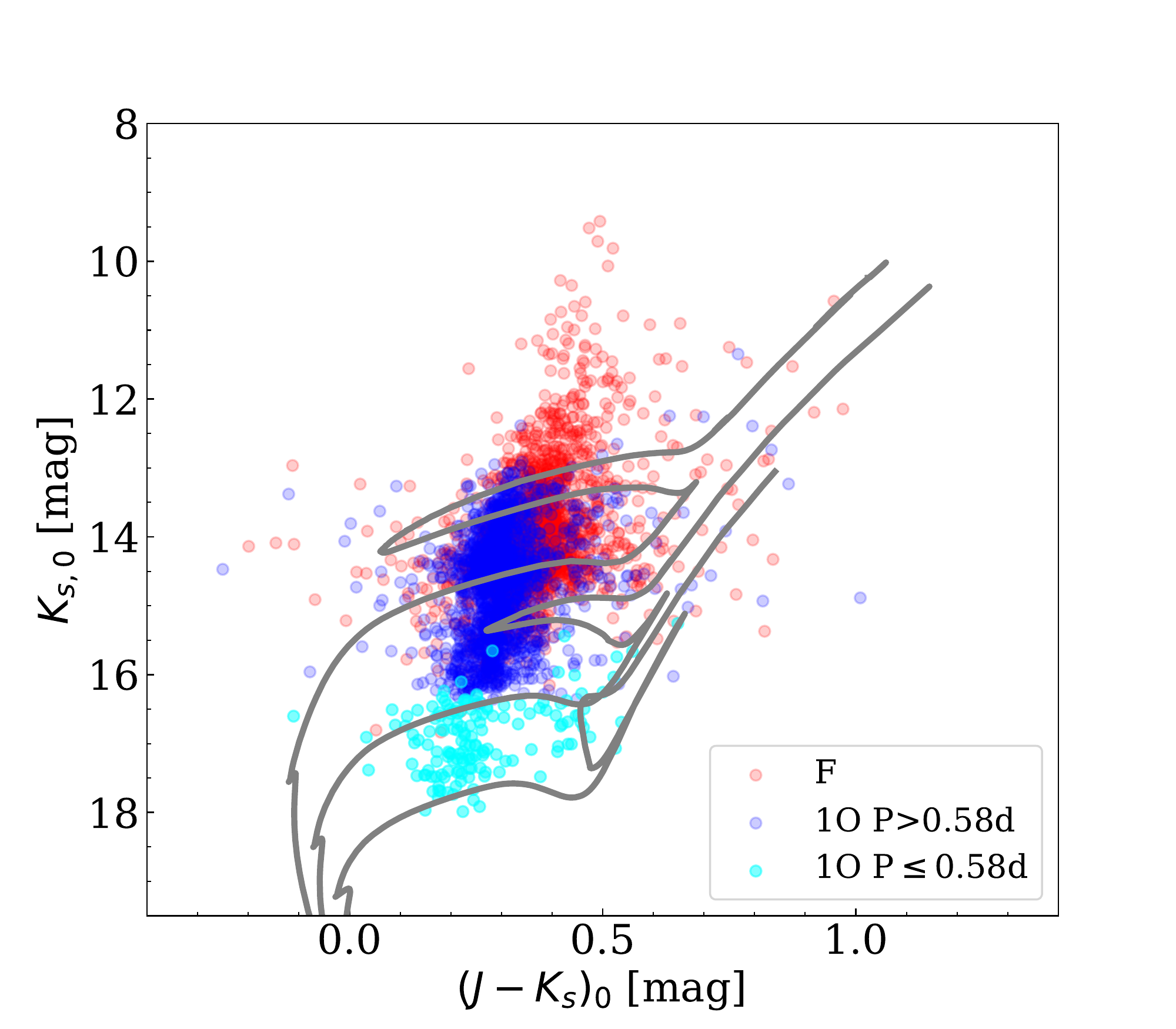}
    \caption{CMD for the investigated DCEPs in the LMC (see labels) in comparison with selected isochrones (grey lines) by \citet{Hidalgo2018} with $Z$=0.008, $Y$=0.257 and ages 70, 220, 520 Myr, from top to bottom.}
    \label{fig:cmdteo}
\end{figure}

To investigate the origin of the break at $P=$0.58 d we found for the 1O pulsators, we displayed  in Fig.~\ref{fig:cmdteo} the Colour--Magnitude Diagram (CMD) of the DCEPs investigated in this paper and compared them with selected isochrones by \citet{Hidalgo2018}, with $Z$=0.008, $Y$=0.257  and ages of 70, 270 and 520 Myr. Most of the DCEPs with $P\leq$0.58 d are confined to a narrower portion of the instability strip (IS) than the longer period 1O DCEPs. As the $PL$ and $PW$ are relations averaged over the width of the IS, this occurrence can produce the break. The comparison with the isochrones suggests also that the pulsators fainter than the break are likely on their first crossing of the IS, an evolutionary stage during which they are still burning hydrogen in a shell while contracting towards the Hayashi track \citep[for details on this subject see e.g.][and references therein]{Anderson2018,Ripepi2021}. Indeed, the so-called blue loop, i.e. the blue ward path in the CMD, characterising intermediate-mass stars burning helium in the core, of the 520 Myr isochrone in Fig.~\ref{fig:cmdteo}, is too short to enter the IS and produce DCEPs. To assess whether or not the first crossing hypothesis is sensible, we would need to check if the number of objects below $P$=0.58 d is compatible with that of the remaining DCEPs, according to the times required to cross the IS in the first and second crossings. This calculation is quite complex and beyond the scope of the present paper. However a rough estimation can be obtained by comparing the time needed by a 3 M$_{\odot}$\footnote{Corresponding to an age of about 265 Myr at the first crossing, see Fig.~\ref{fig:cmdteo} for reference} star to cross the IS at its first crossing with that taken by a 5 M$_{\odot}$\footnote{A DCEP with such a mass has an age $\sim$91 Myr, i.e. representative of the main episode of formation of DCEPs, see Sect.~\ref{sect:ages}} star to cross the IS in the opposite direction during the second crossing. Using tracks taken from the BASTI (a Bag of Stellar Tracks and Isochrones\footnote{http://basti-iac.oa-abruzzo.inaf.it/index.html}) database, which have the same characteristics as the \citet[][]{Hidalgo2018} isochrones, we find that the first crossing takes $\sim$0.5 Myr, while the second crossing takes $\sim$4 Myr, i.e. a ratio of 8 between second and first crossing. To account for the much larger number of second crossing than first crossing DCEPs in our sample, we may simply consider the intervals in magnitude (e.g. in $\kk$) encompassed by first and second crossing DCEPs, respectively, which are of approximately 1.5 and 6 mag (see Fig. ~\ref{fig:figK}), i.e. a ratio of about 4. Therefore, the total ratio between the expected number of DCEPs in their second and first crossing is $\sim 4 \times 8=32$. Since we have about 140 stars with P$<$0.58 d, assuming that all the 1O-mode with P$>$0.58 d and the F-mode DCEPs are in the second and third crossing, we would thus expect a number of DCEPs of $\sim$4500. This is close to the actual number (i.e. $\sim$4700). Therefore, even this very rough calculation seems to support the hypothesis that the break at $P$=0.58 d may be caused by the passage from first to second crossing DCEPs.

\section{Ages}
\label{sect:ages}
An important element of the following analysis is 
the age of the DCEPs, as this quantity allows us to connect 
the pulsational properties of the DCEPs to
 those of the host stellar population. In this way we
can use the location and the 3D structure of the DCEPs as a proxy to study the spatial distribution of the LMC young population (age $\sim$10--500 Myr).

To estimate the ages of the target DCEPs we  
adopted the Period--Age ($PA$) and Period--Age--Colour ($PAC$) relations for F and 1O pulsators at
$Z$=0.008 recently calculated by \citet[][]{Desomma2021} which are an update of previous relations from the same group \citep[][]{Bono2005}. Note that these models do not include rotation, which can produce significantly larger ages with respect to canonical models with and without overshooting \citep[see e.g.][]{Anderson2016}. However, in the following discussion we will use the DCEPs' ages in a relative way, investigating the differential distribution of ages across the LMC, so that the absolute values of the ages are not a major concern. To calculate the  $(V-I)_0$ colour needed in the $PAC$ relations, we used
OGLE\,IV $V,I$ magnitudes and the reddening values listed in  Table~\ref{table:averages}. 
The resulting ages are in the range $\sim$20--250 Myr, with errors of $\sim 10$ per cent and $\sim$6.5 per cent for F and 1O pulsators, respectively. We verified that the $PA$ and the $PAC$ relations provide the same age values within a few Myr \citep[see also][]{Desomma2021}. Therefore, in order not to lose the DCEPs for which OGLE IV does not provide the $V$ magnitude, we decided to use the $PA$ relation for all the pulsators. 

\begin{figure}
	\includegraphics[width=\columnwidth]{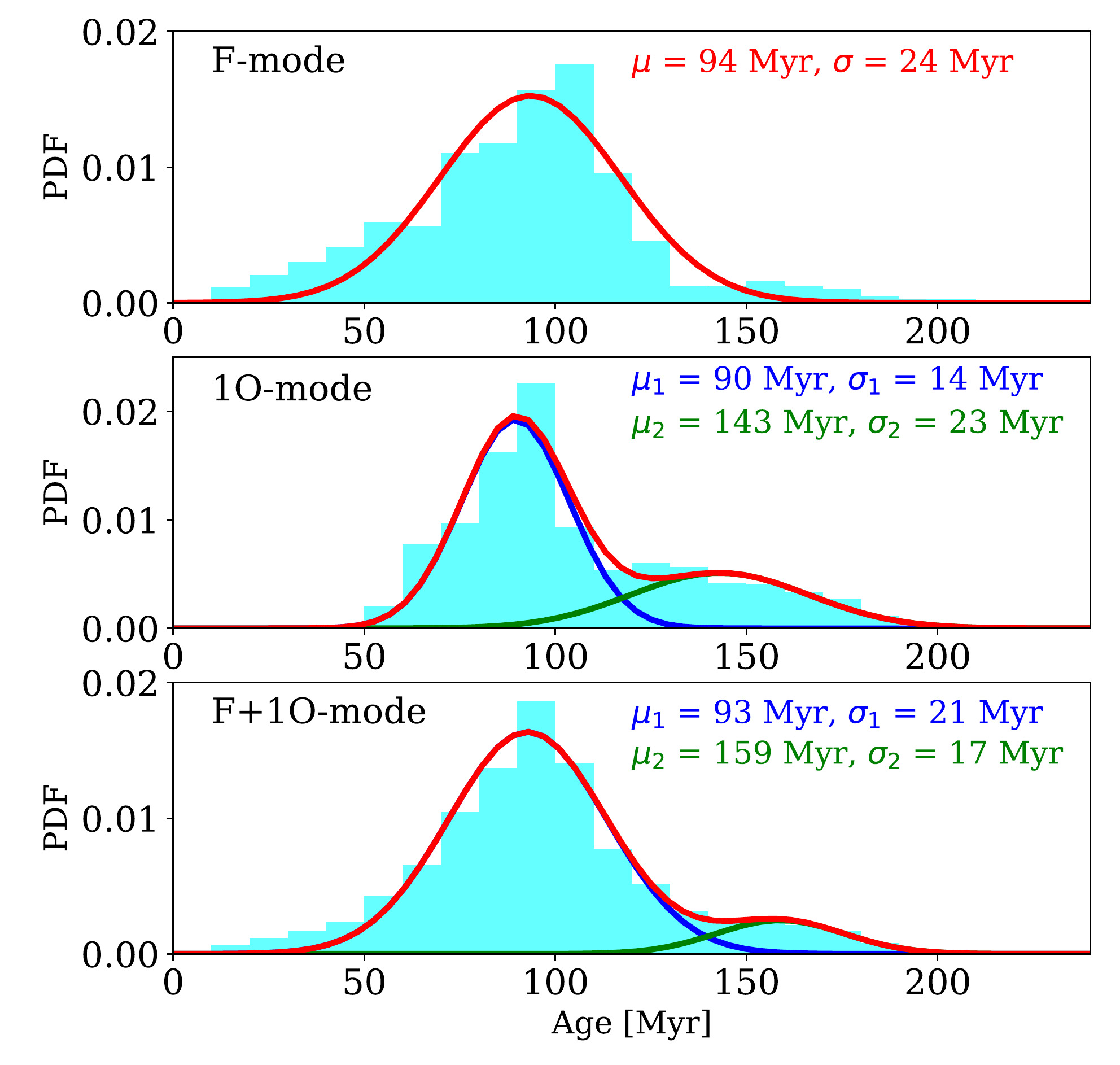}
    \caption{Light blue histograms show the age distribution of the LMC DCEPs divided in F-mode (top panel), 1O-mode (middle) and the full sample (bottom panel). The red line represents the fit to the data, obtained with only one (top panel) or the sum of two (middle and bottom panels) Gaussian functions. In the latter two cases, blue and green lines show the two individual Gaussian functions used to fit the data. Their means and dispersion are labelled with the proper colour.}
    \label{fig:histoAges}
\end{figure}

\begin{figure*}
	\includegraphics[width=17cm]{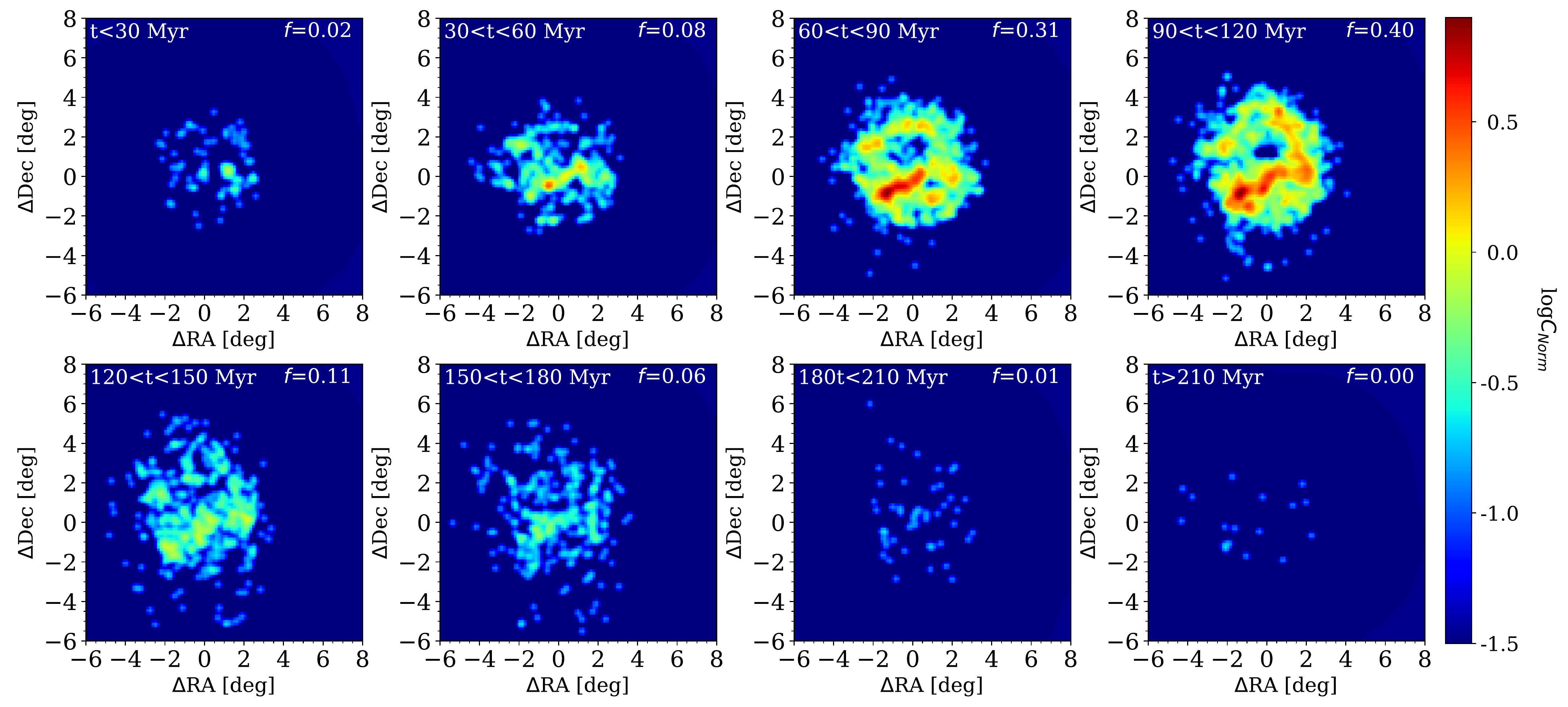}
    \caption{Map of the DCEPs in the LMC for varying age bins (see labels). The numbers on the  top-right of each panel represent the fraction ($f$) of DCEPs in each age interval with respect to the total. The pixel size is $14^{'} \times 14^{'}$ or $\sim 200 {~ \rm pc} \times 200$ pc.}
    \label{fig:mapAges}
\end{figure*}

Figure~\ref{fig:histoAges} top and middle panels show the age distributions of the F- and 1O-mode DCEPs  in the LMC, respectively. The two distributions are somewhat different. Indeed, even if the main peak of both histograms is placed at about 90 Myr, the F-mode pulsators extend to younger ages, a direct consequence of the larger masses (lower ages) spanned by these objects. Conversely, the 1O-mode DCEPs are relatively more numerous at ages younger than about 120 Myr with respect to the F-mode pulsators. More precisely, the F-Mode distribution can be fitted with one Gaussian peaking at 94 Myr with a dispersion of 24 Myr, while the 1O-mode histogram is best-fitted by two Gaussians having means of 90 and 143 Myr and dispersion of 14 and 23 Myr, respectively.
The bottom panel of Fig.~\ref{fig:histoAges} shows the cumulative age histogram including both F- and 1O-mode DCEPs. The distribution can be well modelled with two Gaussians having means of 93 and 159 Myr with dispersions of 21 and 17 Myr, respectively. We can roughly assign the value of 2$\sigma$, i.e. $\sim$40  Myr as the time-scale of both star formation episodes that produced DCEPs in the LMC. Therefore, according to our results, the DCEPs in the LMC were formed in two main episodes of star formation lasting $\sim$40 Myr which  happened 93 and 159 Myr ago. The first event was less efficient than the second, that is responsible for the formation of more than 80 per cent of the DCEPs in the LMC.  

Note that the histograms of Fig.~\ref{fig:histoAges} do not include the DCEPs with $P<0.58$ days, as the $PA$/$PAC$ relations are not useful for pulsators in this period range. These objects are certainly older than the bulk of DCEPs, as shown in Fig.~\ref{fig:cmdteo}, where the isochrones overlapping with the very low period DCEPs have 220 Myr$<$t$<$520 Myr.

The high precision of the DCEP age determinations allows us to trace spatially the onset of the star formation for the young population in the LMC of which the DCEPs are a proxy. This is shown in Fig.~\ref{fig:mapAges}, where we plot the distribution on the sky of the LMC DCEPs at intervals of 30 Myr in age. Before proceeding with the discussion, in the following we will refer to different substructures of the LMC which for the most part have already been described in the literature. For the benefit of the reader, we display in Fig.~\ref{fig:mapWrittenSubstructures} the substructures of the LMC that will be mentioned in this paper. 
Figure~\ref{fig:mapAges} shows that the youngest DCEPs have a clumpy distribution with no clear membership to any major feature, except for the Western Bar (WB), the DCEPs in this age interval are only2 per cent of the total. In the age interval $30<t<$60 Myr we find clumps of DCEPs in the bar and in the Northern Arm 1 (NA1), but the total number of DCEPs is still modest, i.e. 8 per cent. In the age interval $60<t<$90 Myr the bulk formation of DCEPs is visible in almost all of the galaxy (31 per cent of the total DCEPs), with the exception of the WB and of the Northern Arm 2 (NA2). In the successive age interval $90<t<120$ Myr the DCEPs formation is active almost everywhere in the LMC, but in particular in the whole bar, in the NA2 arm and in the extreme portion of the NA1. The sub-structuring of the bar is also noteworthy. For example, a spot of DCEPs appears at coordinates ($-$1, $-$1.5) (deg), with a dimension of about 400 pc (we call this feature the South Eastern Structure or SES). At later ages the DCEPs formation is less strong and more diffused, even if the majority of the objects are formed along the bar. At an age of 180 Myr and older only a small number of DCEPs were formed. These results are in substantial agreement with those shown by \citet{Jacy2016}.

The maps displayed in Fig.~\ref{fig:mapAges} can be compared with those based on the extended SFH study by \citet[][their fig. 5]{Mazzi2021}. The general morphology of the two set of maps appears in agreement. As an example, in both studies the formation of the NA2 is posterior to that of the NA1, even if the absolute ages at which these events occurred are different in our and \citet{Mazzi2021}'s work, owing to the different sets of evolutionary models employed.

\section{3D geometry of the LMC}

To determine the three-dimensional structure of the young population of the LMC as traced by the DCEPs we have first to estimate the relative distances of each DCEP with respect to the centre of the LMC. 
There are several determinations of the LMC centre, with considerable differences between the different estimates \citep[see e.g.][and references therein]{Luri2021}, depending on the adopted method. To be consistent with the results by \citet{Luri2021} that will be used in the following, we adopted their same centre, i.e. (81.28, $-$69.78) deg (J2000), according to \citet{vanderMarel2001}. This value is however not far away from the centroid of the DCEP distribution (80.68, $-$69.24) deg (J2000). 

The next step is to estimate the relative distances of each DCEP from the adopted LMC centre. 
We used only the $PL$ and $PW$ relations in the $V$, $J$, $\kk$ bands because the relations involving the $Y$-band have a lower accuracy. For each DCEP we then calculated the magnitude difference with respect to the $PL$/$PW$ relations:
\begin{equation}
\Delta Mag =  Mag-(a \times {\rm log}P+b)
\end{equation}
\noindent 
where $Mag$ can be $J$/$\kk$ or $W(V,K_\mathrm{s})$/$W(J,K_\mathrm{s})$ and the $a$ and
$b$ values are those listed in Table~\ref{table:pl}. To transform these
$\Delta Mag$ values into absolute distances, we adopted the distance modulus of
the LMC by \citet{degrijs2014} $\mu=18.49$ mag 
corresponding to $D_{\rm LMC}=49.9$ kpc. However, the precise
value of $\mu$ does not affect the subsequent analysis
\citep[see e.g.][]{Subramanian2015}. The individual distances are
hence calculated as:
\begin{equation}
D_i = D_{\rm LMC} 10^{(\Delta Mag_i/5)}
\label{eqn:distance}
\end{equation}

\noindent
To estimate the errors on the individual relative distances, we first 
consider the photometric errors
 that are shown in Fig.~\ref{fig:errors}. It can be seen that the large majority of the
 DCEPs show errors less than 0.02 mag and 0.01 mag in the $J$ and $K_\mathrm{s}$ bands, respectively. 
 At the distance of the LMC, these values correspond to errors of $\sim$0.4 kpc and 0.2 kpc, respectively. 
 As for the $PW$ relations, we propagated the errors obtaining a relative uncertainty of $\sim$0.45 kpc for all combinations of colours used in this work \citep[an
uncertainty of 0.02 mag for the OGLE\,IV  Johnson-$V$ band was adopted here,
see][]{Jacy2016}.  However, the intrinsic dispersion of the
adopted $PL$/$PW$ relations due to e.g. the finite width of
the IS, mass-loss, rotation, etc, represents the major source of error. 
The finite width of the IS impacts mainly on the 
$PL$ relations, although in the NIR bands the effect is
reduced. Even better are the $PW$ relations, indeed for the $PW(J,K_\mathrm{s})$ and
$PW(V,K_\mathrm{s})$ relations, the intrinsic dispersion is estimated to be 
 $\sim$0.04 and 0.05 mag, respectively \citep{Inno2016}. Hence, summing
up in quadrature all the sources of errors, our 
best individual relative distance errors are $\sim$1.0 and 1.2 kpc 
for the $PW(J,K_\mathrm{s})$ and $PW(V,K_\mathrm{s})$, respectively.

\subsection{Viewing angles}
\label{Sect:cartesian}

The individual distances can be used to calculate the Cartesian coordinates for the LMC DCEPs. 
Adopting the usual formalism by \citet{vandermarelCioni2001} and \citet{Weinberg2001} we have:
\begin{eqnarray}
X_k &=& -D_k\sin(\alpha_k - \alpha_0)\cos\delta_k  \nonumber \\
Y_k &= &D_k\sin\delta_k\cos\delta_0 - D_k\sin\delta_0\cos(\alpha_k -
\alpha_0)\cos\delta_k  \nonumber \\
Z_k &= &D_0-D_k\sin\delta_k\sin\delta_0 - D_k\cos\delta_0\cos(\alpha_k -
\alpha_0)\cos\delta_k , \nonumber 
\label{eq:cartesian}
\end{eqnarray}

\noindent 
where $D_0$ is the mean distance of the LMC, $D_k$ is the distance to each DCEP calculated using   
Eq.~\ref{eqn:distance}, 
($\alpha_k$, $\delta_k$) and ($\alpha_0$, $\delta_0$) represent the RA and Dec of 
each DCEPs and the centre of LMC, respectively. By definition, 
the X-axis is anti-parallel to the RA axis, the Y-axis is parallel to the declination axis, and the Z-axis has its origin in the centre of the LMC.

The Cartesian coordinates can be used to fit the distribution of the LMC DCEPs with a plane, assuming that a planar distribution can describe the location of the pulsators along the LMC. This is a crude but necessary approximation, as it allows us to discuss the LMC geometry on a quantitative basis. In practice we fitted the following equation to the data: 
\begin{equation}
Z_{j,k}=A_j X_{j,k}+B_j Y_{j,k}+C_j     
\label{eq:plane}
\end{equation}

\noindent
where $k\in1,N$ and $N$ in the number of DCEPs used, while the index $j$ refers to the different fits to the data  carried out for all the cases summarised in Table~\ref{table:pl}, except for those involving the $Y$ band.

The viewing angles of the LMC disc, i.e. the inclination $i$ and the position angle of the line of nodes $\theta$ \citep[see e.g.][]{vandermarelCioni2001} can be calculated from the coefficients of Eq.~\ref{eq:plane} as follows:
\begin{eqnarray}
i_j= \arccos \left( \frac{1}{\sqrt{A_j^2+B_j^2+1}}\right) \\
\theta_j=\arctan \left( -\frac{A_j}{B_j}\right) + sign (B_j) \frac{\pi}{2}, 
\label{eq:ang}
\end{eqnarray}

\noindent
where the $j$ index has the same meaning as before. 
To estimate the coefficients of the plane, we used the python package {\tt LtsFit} \citep{Cappellari2013} which provides robust results and performs a reliable outlier removal. The resulting $i$ and $\theta$ values are listed in Table~\ref{table:inclination} together with their errors, which were calculated using standard error propagation rules. 

An inspection of Table~\ref{table:inclination} reveals that the inclination value is rather insensitive to the sample (F, 1O or F+1O pulsators\footnote{Here and in Table~\ref{table:inclination}, with F+1O we indicate the sum of the F and 1O pulsator samples}) or to the different PL/PW relations adopted in its calculation. The only clear differences are in the associated errors. Indeed, as expected, the adoption of the full sample (F+10 DCEPs) yields smaller errors. The use of the tightest relations $PL\kk$, $PWJ\kk$ and $PWV\kk$ has a similar effect. On the other hand, the $\theta$ value appears to depend on the sample adopted to calculate its value. Indeed, the F  DCEPs  provide systematically larger $\theta$ values than the 1O, even if values agree within two $\sigma$. 
This is consistent with the different 2D distribution of F and 1O DCEPs shown in Fig.~\ref{fig:difference}. The figure shows that the distributions of F- and 1O-Mode DCEPs are different. The F-mode pulsators have a roundish and concentrated distribution, with the stars clearly tracing the spiral arms and the bar in all their extensions. The 1O DCEPs occupy a significantly larger area and their distribution is less symmetric and ordered than the F pulsators, as they are placed also outside the major features, spiral arms and bar. The lack of 1O DCEPs in WB and at the eastern tip of the bar is noteworthy. A possible explanation of this different distribution rests in the different SFHs between the two pulsation modes, as the 1O DCEPs have on average smaller masses than F pulsators and consequently are not present in star forming regions younger than 50--60 Myr (see Fig.~\ref{fig:mapAges}). 
The different spatial distributions can therefore be responsible for the different $\theta$ values obtained with the two samples. The use of the combined sample provides intermediate $\theta$ values, with much reduced errors, too. As the inclination of the disc described by the whole sample is almost identical to that of the individual sub-samples, we judge that it is correct to retain as best values for $\theta$ those obtained with the whole sample. For both $\theta$ and $i$, we  therefore adopt the weighted mean of the values obtained with the $PL\kk$, $PWJ\kk$ and $PWV\kk$ relations and the full DCEP sample as our best estimate for these parameters, i.e. $\theta$=145.6$\pm$1.0 deg and $i$=25.7$\pm$0.4 deg.        

We can now compare our results with those coming from similar studies using DCEPs as stellar tracers (see Table~\ref{tab:litComparison}). The inclination values found by the most recent DCEP-based  studies of \citet{Jacy2016}, \citet{Inno2016} and \citet{Deb2018} are in very good agreement with ours. Older works by \citet{Nikolaev2004}  and \citet{Haschke2012} provided significantly larger values, but this is a clear consequence of the much smaller sample of DCEPs adopted in those studies. More interesting is the analysis of the $\theta$ parameter. Indeed, all the previous works using DCEPs, except \citet{Haschke2012} report values around 150 deg, i.e. values close to what we find with the F-mode pulsators only. Moreover, both \citet{Inno2016} and \citet{Deb2018} do not find differences between $\theta$ values obtained with F- and 1O-mode DCEPs. The explanation for this apparent discrepancy is the different sample of objects used in the calculation. Our data allows us to use also the fainter 1O-mode DCEPs that, even with the exclusion of $P<0.58$ d objects, are significantly more numerous than in those previous works. As discussed above, the spatial distribution of the 1O DCEPs is more extended than for the F pulsators. Therefore it is reasonable to conclude that the difference between our best estimate of $\theta$ and previous works, of more than 3$\sigma$, can be ascribed to the mere difference in the spatial displacement of the samples used in the calculation. 

\subsection{Viewing angles of the LMC bar and disc at different radii}

\begin{figure}
	\includegraphics[width=8.7cm]{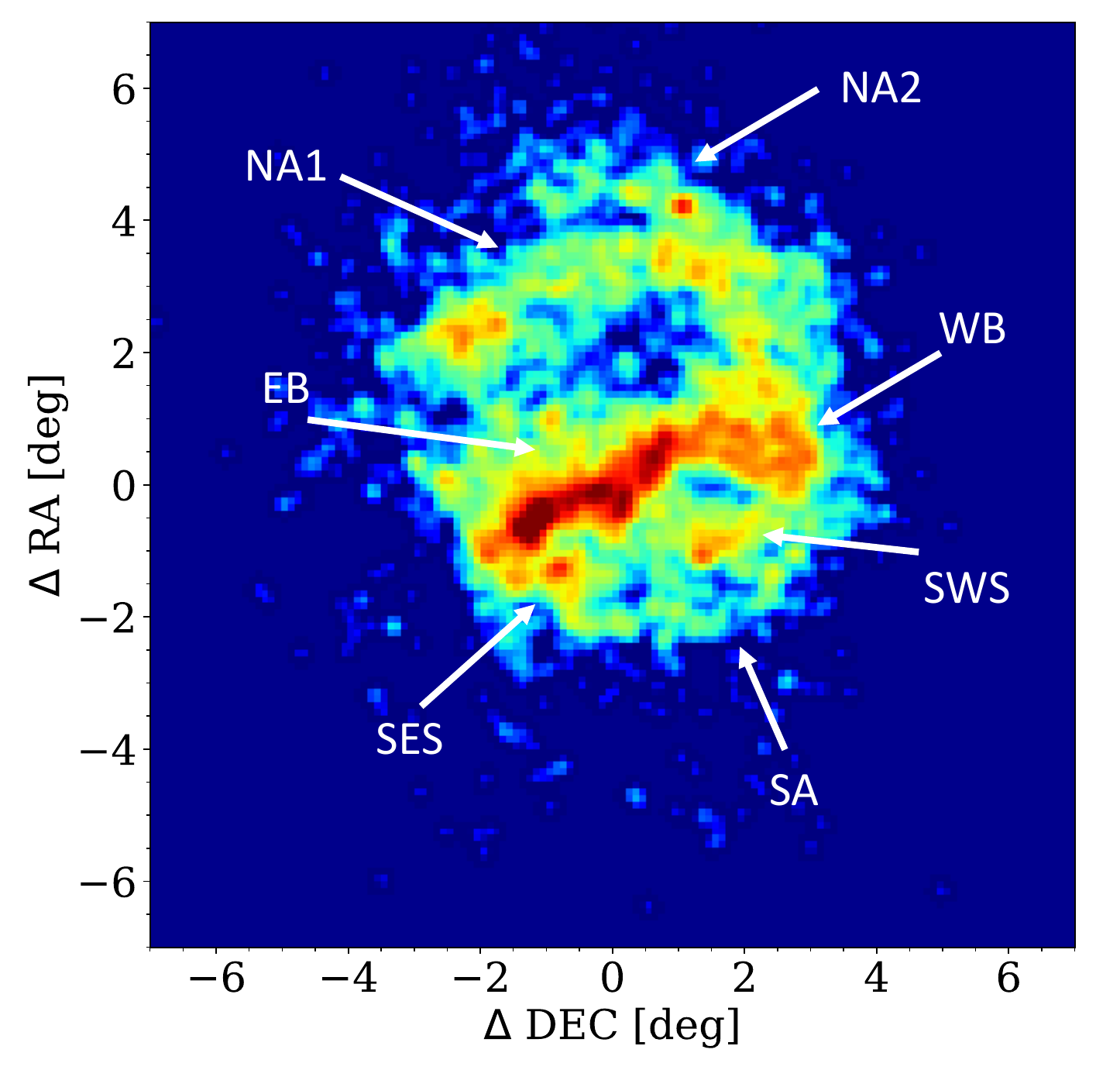}
    \caption{Map of the DCEPs in the LMC. The major substructures discussed in this paper are shown. The meaning of the acronyms is the following: NA1 = Northern Arm 1; NA2 = Northern Arm 2; EB = Eastern Bar; WB = Western Bar; SES = South Eastern Structure; SWS = South Western Structure; SA = Southern Arm.}
    \label{fig:mapWrittenSubstructures}
\end{figure}

The analysis of the viewing angles of the LMC is concluded by testing whether they vary with the distance to the LMC centre. We also investigate if bar and disc viewing angles are compatible or show significant discrepancies.
The estimate of the viewing angles including/excluding the bar sample was carried out for $PW(J,\kk)$ only, as the other combinations of filters/colours give the same results. The bar is defined inside a polygon with the following edges in RA and Dec: (88.90, $-$70.45); (76.30, $-$68.01); (74.28, $-$69.03); (87.95, $-$71.49) (deg) including 1385 DCEPs covering essentially the Eastern Bar (EB, see Fig.~\ref{fig:mapWrittenSubstructures}), whereas the disc sample includes 2838 stars. 
The result of this investigation is reported in Fig.~\ref{fig:inclinationBar}. Using the whole sample of DCEPs, both $i$ and $\theta$ vary considerably within LMC galactocentric distances smaller than 4--5 kpc, highlighting the importance of the adoption of an extended sample in deriving the viewing angles. The high degree of variation is also due to the preponderance of the bar sample at low galactocentric radii. Indeed, if we calculate the viewing angles separately for the bar and disc samples as defined above, we find $i$=37.4$\pm$4.7 deg, $\theta$=131.4$\pm$9.2 deg and $i$=26.4$\pm$0.7 deg, $\theta$=141.6$\pm$1.9 deg, respectively (see also Fig.~\ref{fig:inclinationBar}). Therefore, the bar appears to have a significantly different inclination with respect to the disc, which however dominates when the two samples are used together, indeed in this case the inclination $i$=25.25$\pm$0.05 is more similar to that of the disc sample than to the bar one. This result is in perfect agreement with previous literature findings \citep[e.g.][]{Nikolaev2004,Inno2016,Niederhofer2021}.    

\begin{figure}
	\includegraphics[width=9cm]{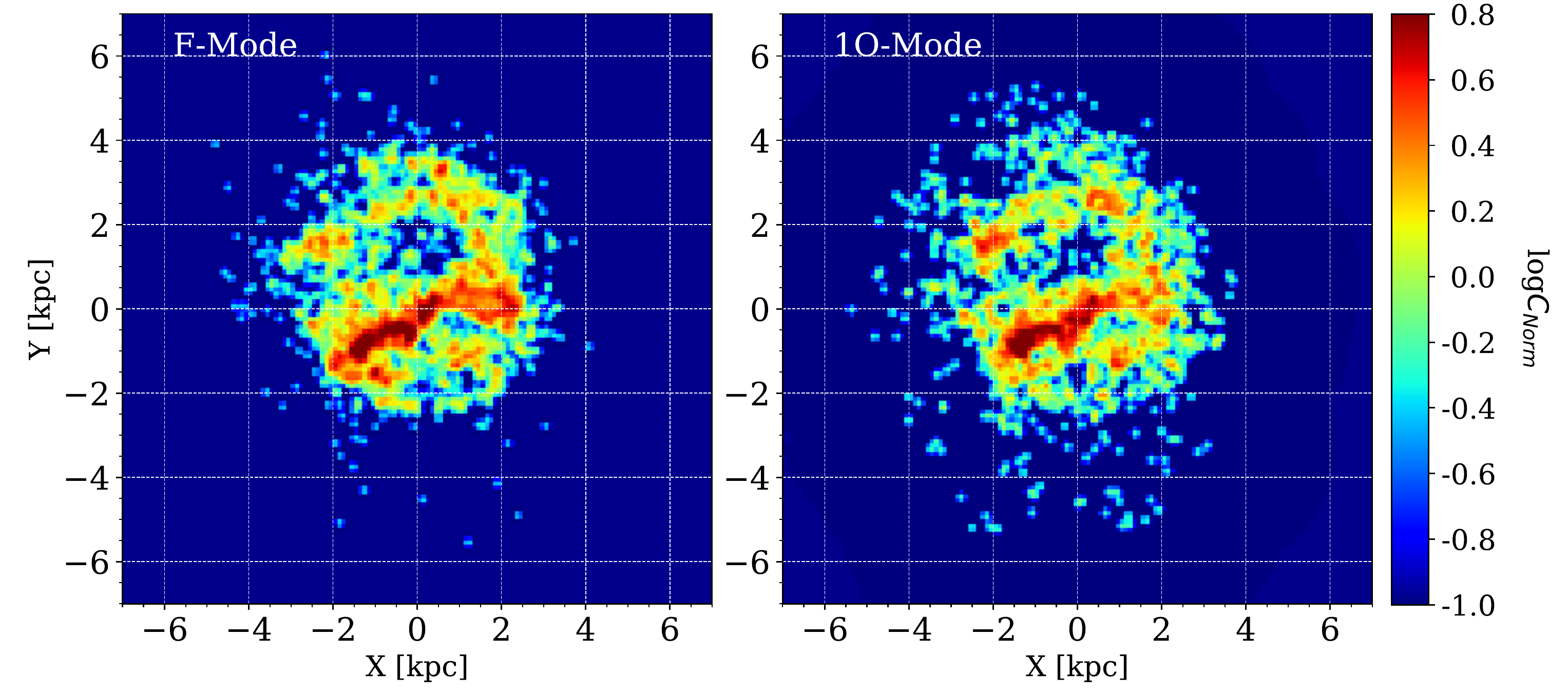}
    \caption{The left and right panels show the distribution in the XY plane of the F and 1O LMC DCEPs, respectively. The pixel-size is 50 pc $\times$ 50 pc. Each 2D distribution was smoothed by means of a KDE (kernel density estimate) algorithm with bandwidth=80 pc.}
    \label{fig:difference}
\end{figure}

\begin{figure}
	\includegraphics[width=\columnwidth]{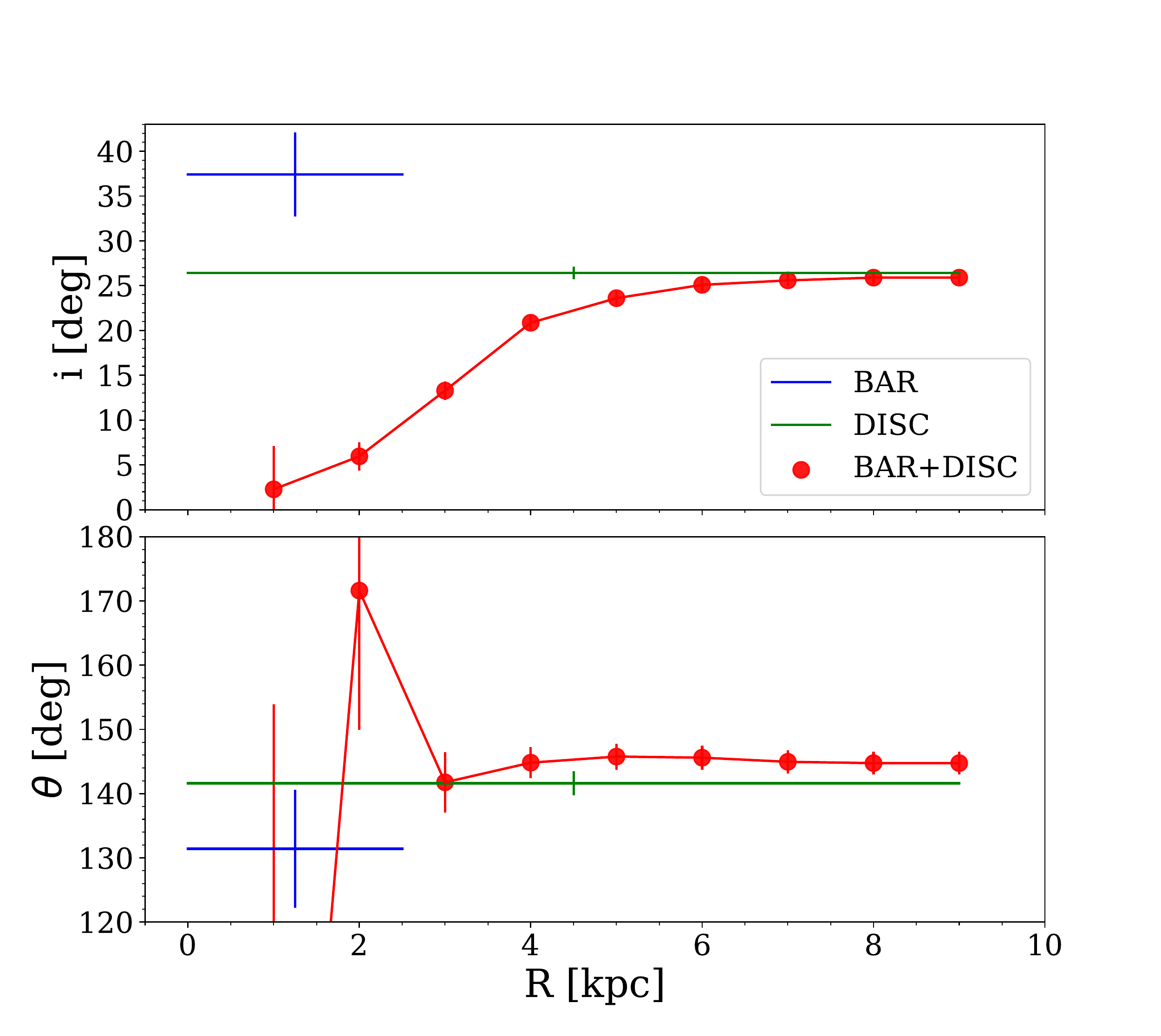}
    \caption{Inclination (top panel) and position angle of the line of nodes (bottom panel) at varying radii for the whole LMC DCEP sample (red filled circles connected by a red line). The green and blue lines display the values obtained for the separated disc and bar samples, respectively. In all the cases the length of the line represents the approximate radius of the sample used for the calculation, while the vertical lines give the uncertainty on the measure. The error bar is conventionally placed in the middle of the horizontal blue and green lines.}
    \label{fig:inclinationBar}
\end{figure}

The previous findings show that the LMC bar is not only off-centre with respect to the centre of the disc \citep[see e.g. Fig.~\ref{fig:difference} or][and references therein]{vanderMarel2001,Nikolaev2004,Bekki2009}, but also off-plane. The LMC bar has been simulated by \citet{Bekki2009}, who found that the off-centre feature can be explained by the collision of the LMC (with the bar already formed) with a galaxy (or dark halo) having a mass of a few  per cent of that of the LMC. However, this relatively minor collision cannot explain the off-plane feature, which could however be possibly due to a significantly more violent collision with the SMC. Of course, this possibility should be investigated on the basis of $N$-body modelling of LMC--SMC--MW interaction. In this respect, the results presented here can provide helpful constraints on the pericentric distance and the mass of the SMC during the last LMC--SMC interaction about 150 Myr ago \citep[see e.g.][]{Cullinane2021}.

\begin{figure*}
	\includegraphics[width=14cm]{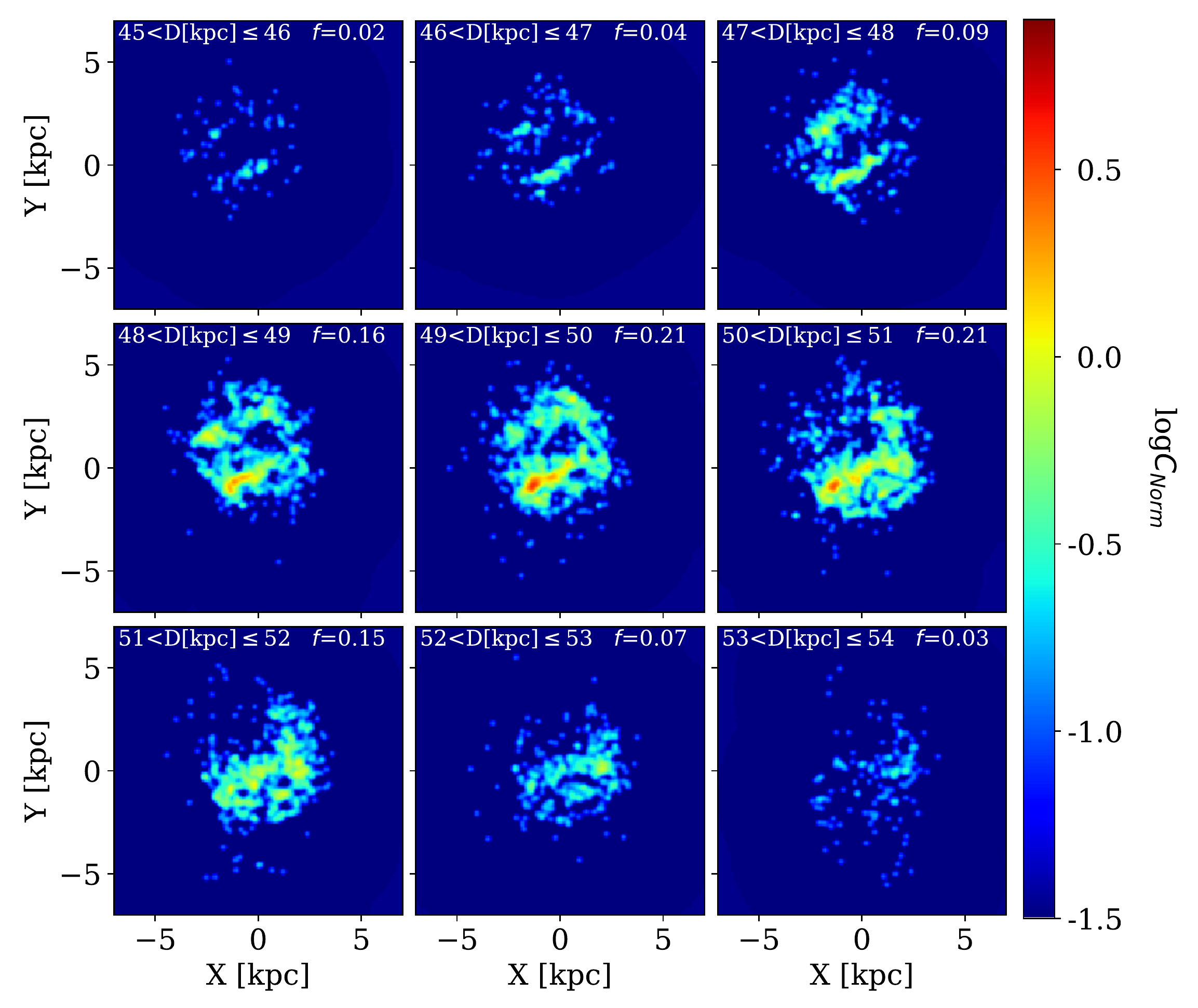}
    \caption{Distribution of DCEPs in the LMC at varying distances from the Sun. The pixel-size is 100 pc x 100 pc. The numbers on the top-right of each panel represent the fraction ($f$) of DCEPs in each distance interval with respect to the total. In each panel the 2D distributions were smoothed by means of a KDE algorithm with bandwidth=120 pc.}
    \label{fig:mapXYZ}
\end{figure*}

\begin{table*}
\caption{Inclination and position angle values based on DCEPs from different literature sources. Note that the errors listed for the values by \citet{Inno2016} are those recalculated by \citet{Deb2018}.}
\footnotesize\setlength{\tabcolsep}{3pt}
\label{tab:litComparison}
\begin{center}
\begin{tabular}{lccc}
\hline 
\noalign{\smallskip} 
Inclination    &  Pos. Angle & PL/PW  & source  \\
   (deg)    &  (deg) &    \\
\noalign{\smallskip} 
\noalign{\smallskip} 
\hline 
154.7$\pm$1.4 &  25.1$\pm$0.4    &  $PL(V,I,J,H,\kk,[3.6],[4.5])$   & \citet{Deb2018}      \\ 
150.80$\pm$1.15 & 25.05$\pm$1.15  &  $PW(V,I),PW(H,J,\kk)$	& \citet{Inno2016}  \\
151.4$\pm$1.5 &  24.2$\pm$0.6   &  $PW(V,I)$	&   \citet{Jacy2016}   \\
150.2$\pm$2.4 &  31$\pm$1       &  $PL(V, R, J, H,K)$	& \citet{Nikolaev2004}   \\
116$\pm$18 &    32 $\pm$4        &  $PL(V,I)$	& \citet{Haschke2012}     \\
\noalign{\smallskip}
\hline
\end{tabular}
\end{center}
\end{table*}

\section{Sub-structures in the LMC}

The distribution of the DCEPs in the LMC at varying distances from the Sun is presented in Fig.~\ref{fig:mapXYZ}. The different panels of the figure show the high structuring of the LMC: the NA1 arm and the EB  are closer to us by some 2--3 kpc with respect to the centre of the galaxy. Conversely the South-West portion of the LMC is 2--4 kpc farther. These results are in agreement with all previous works by \citet{Nikolaev2004,Inno2016,Jacy2016,Deb2018}.
As for the bar, it is visible in almost all the panels, revealing that its thickness is larger than $\sim$4-5 kpc.

\begin{figure*}
	\includegraphics[width=14cm]{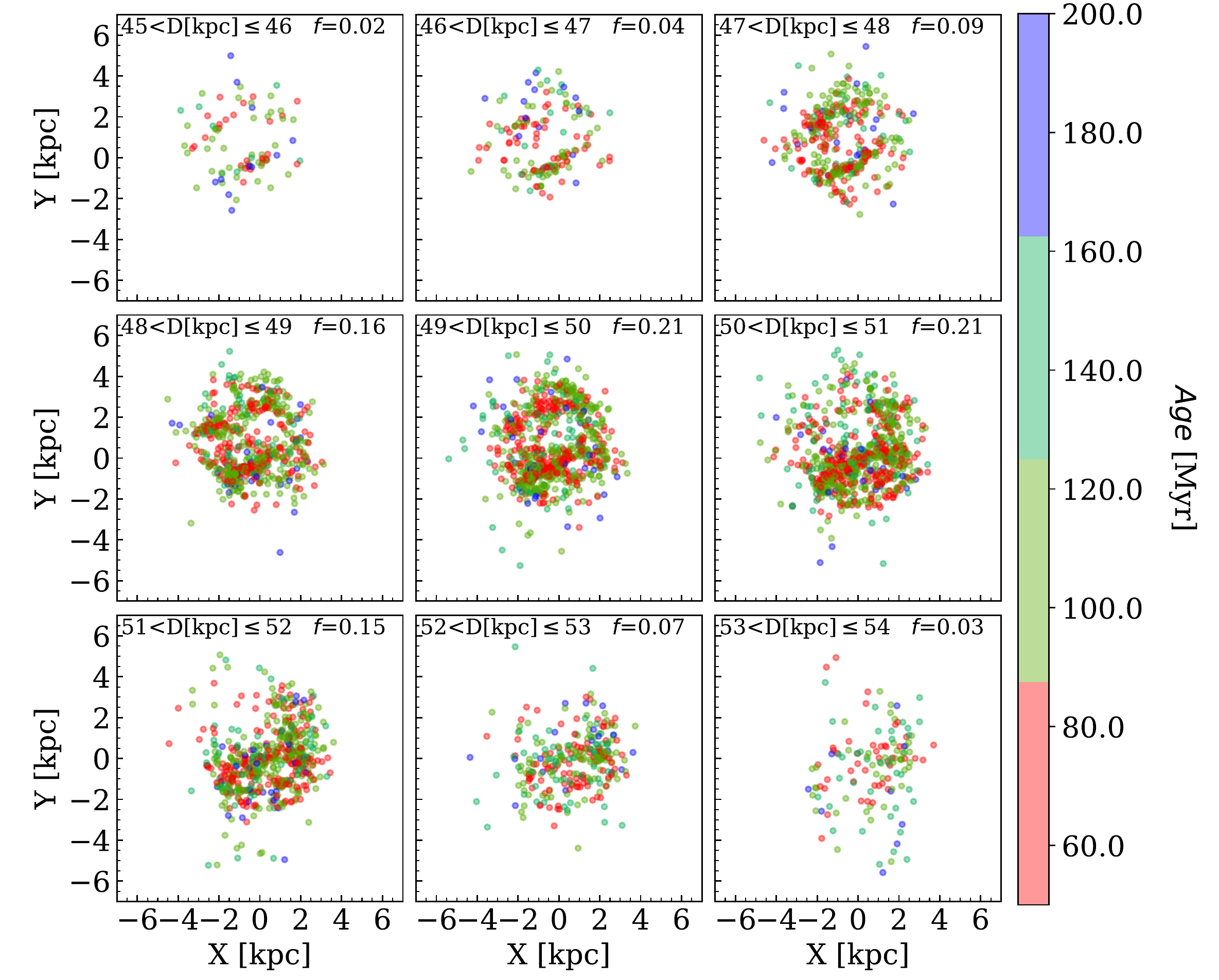}
    \caption{Distribution of DCEPs in the LMC at varying distances from the Sun, subdivided into four age bins (see colour-bar on the right).}
    \label{fig:mapXYZ_age}
\end{figure*}

It is interesting to add the age information to Fig.~\ref{fig:mapXYZ} in order to search for possible space and temporal correlations. This is shown in Fig.~\ref{fig:mapXYZ_age}. An analysis of the figure reveals that NA2 shows the oldest stars to be at close distances while the SA and SWS structures are similarly populated but at large distances. Moreover, in general the younger stars tend to clump more than older ones. Young DCEPs also appear to be more centrally concentrated as they compete in number with "middle" aged DCEPs between 48 and 51 kpc and especially in the 49--50 kpc bin. At closer and farther distances older DCEPs are dominant. It is also remarkable that NA1 and NA2, even though placed in the same distance bin (49--50 kpc), show completely different ages, with NA1 clearly younger than NA2, as already found in Sect.~\ref{sect:ages}. Therefore, at least in this region, the DCEP formation proceeded from west to east.  
According to the $N$-body models by \citet{Cullinane2021} the last pericentric passage of the SMC happened about 150 Myr ago. This close encounter could be responsible for the peak of DCEP formation in the LMC, as well as for the most recent episode in the SMC that dates to $\sim$120 Myr ago \citep[the other DCEP peak in the SMC is estimated at $\sim$220 Myr ago, see e.g.][and references therein]{Ripepi2017}. We wonder if there is a spatial signature of this chain of events in the present age--spatial distribution of the LMC DCEPs. According to \citet[][]{Cullinane2021} the signature of the last interactions between LMC and SMC should be visible in the south-west region of the LMC. Looking at the last bin in distance (53--54 kpc) it is possible to see that the DCEPs are located  mainly on the west side and they do not show any particular shape (disc, bar or spiral arm). The ages of the DCEPs appear mixed, apart from the very south west, where only old stars can be found. This DCEP distribution could be compatible with the disturbance caused by the last SMC passage, but it is difficult to firmly prove this connection.

\begin{figure}
		\includegraphics[width=9cm]{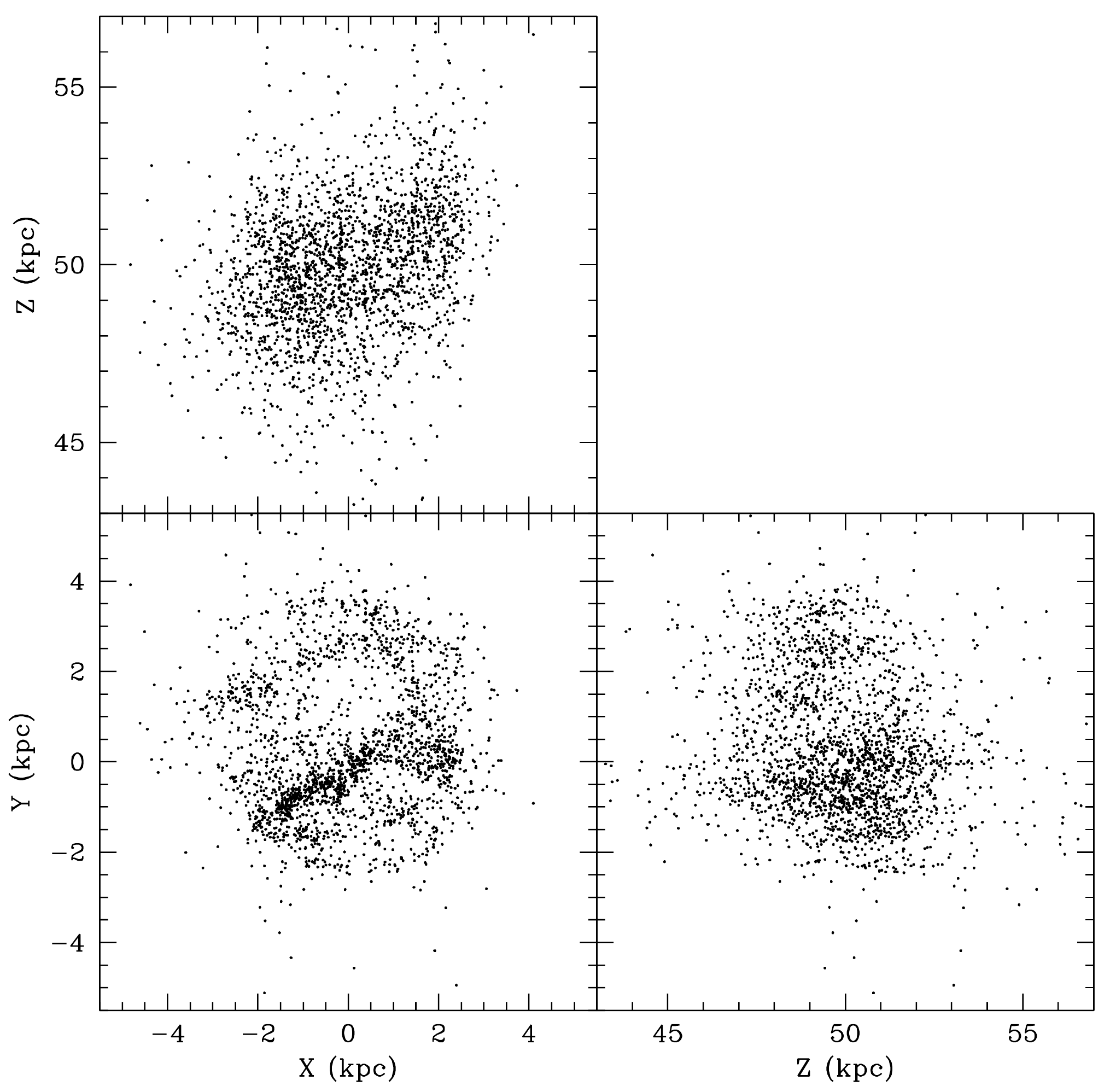}
\caption{Projection of the distribution of LMC DCEPs in the XY,YZ,XZ planes.}
    \label{fig:XYZ_3panels}
\end{figure}

More details about the 3D distribution can be inferred from the analysis of the various projections in 2D, as shown in Fig.~\ref{fig:XYZ_3panels} in the case of the $PWJ\kk$ relation that we will use hereinafter for the analysis because it provides, together with the $PWV\kk$ relation, the tightest relationships, and consequently, the most precise individual distances. We verified that the results do not vary significantly using the $PWV\kk$. 

It is particularly interesting to analyse the XZ and YZ projections shown in Fig.~\ref{fig:XYZ_3panels} from which it is possible to appreciate the high structuring of the LMC disc/bar. From the XZ projection the disc/bar appear to have a "W" shape 
instead of a planar structure with the easternmost stem of the "W" less prominent than the westernmost one (at coordinates 2, 52 kpc). The latter can be interpreted as the warp of the disc identified by \citet{Nikolaev2004}, even if these authors found a symmetric warp, contrarily to what we see here. The central part of the "W" is difficult to identify from this projection but would correspond in large part to the bar, which, as we shall see below, does not lie on a unique plane. This occurrence can be verified from the YZ projection that shows how the EB and the WB are clearly displaced by more than 1 kpc. 
In general, the YZ projection shows that the profile of the LMC, as depicted by DCEPs, hardly resembles a classic thin disc and rather shows an amorphous shape with  radius $\sim$3 kpc and height ~$\sim$ 5 kpc. Interestingly, \citet{Deb2018} were also able to describe the DCEP distribution in the LMC with a triaxial ellipsoid with two axes very similar to each other and a third axis that is about half the other two.

\begin{figure}
	\includegraphics[width=9cm]{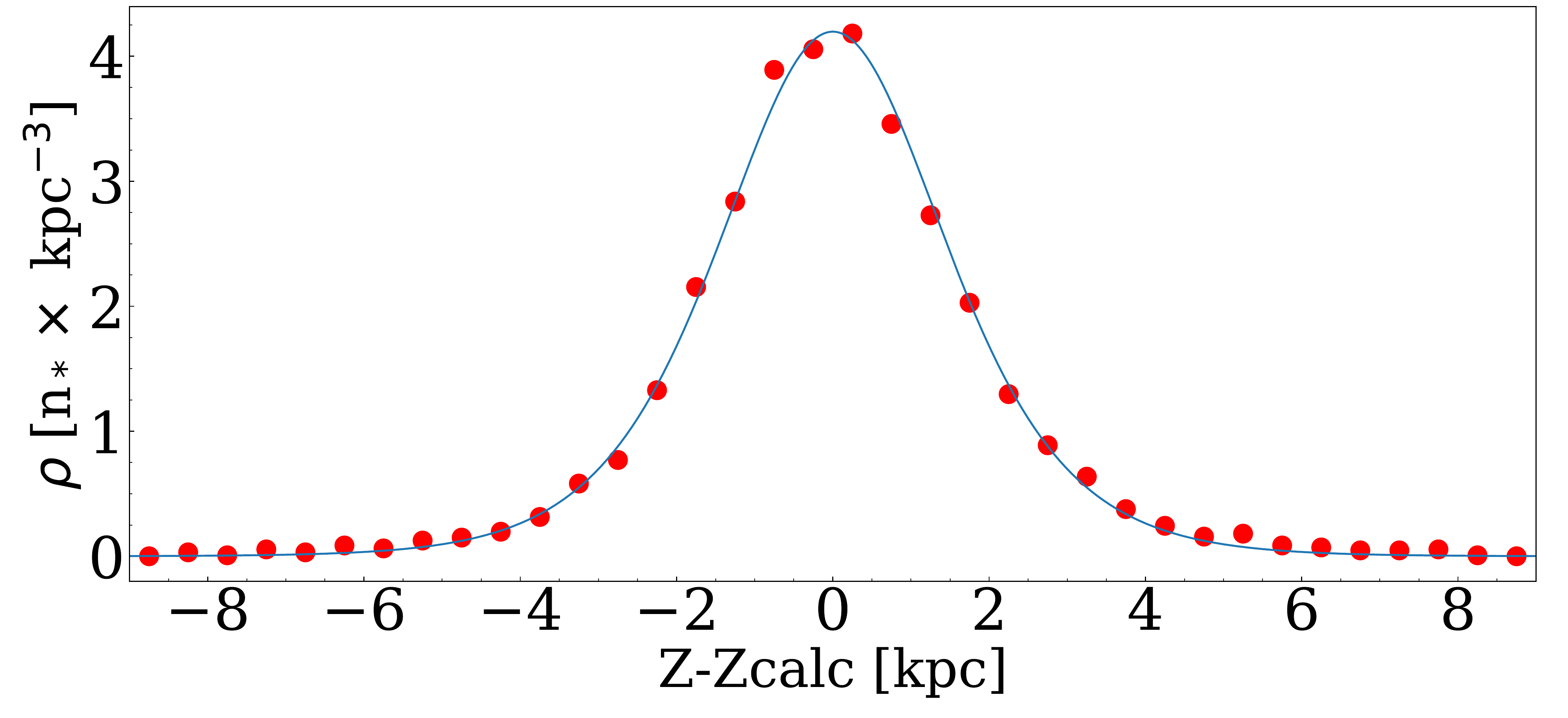}
    \caption{Star density as a function of the height above/below the LMC plane (red filled circles). The uncertainties on the data are smaller than the dots. The blue solid line represents the fit to the data obtained with Eq.~\ref{eq:discHeight}.}
    \label{fig:scaleHeight}
\end{figure}

\subsection{Scale height of the LMC disc}

To further investigate the stellar distribution in the LMC disc, we can remove the plane fitted to the data and analyse the residuals $\Delta Z=Z-Z_{\rm calc}$, where $Z_{\rm calc}$ is the best fitting plane given from the following equation: $Z_{\rm calc}= -(0.396\pm0.016) X  +(0.281\pm0.015) Y -(0.147\pm0.025)$. 

Then, we calculate the stellar density profile along the
Z-direction. The stellar density was calculated considering all the DCEPs in a radius of 9 kpc  (encompassing the large majority of DCEPs).
The resulting stellar densities as a function of the height above/below the plane are  displayed in Fig.~\ref{fig:scaleHeight} (red filled circles). The figure confirms the asymmetry and the high structuring of the LMC, especially for $\Delta Z$ close to 0, where the bar sub-structuring visible in Fig.~\ref{fig:mapXYZ} is  distorting the stellar density profile at values $\sim\pm1.5$ kpc in $\Delta Z$. We have anyway tried to model the stellar distribution of the disc using the functional form derived by  \citet{vanderKruit1981a,vanderKruit1981b,vanderKruit1982,degrijs1997}: 

\begin{equation}
    \rho^*=\rho^*_0 {\rm sech}^2(\Delta Z/ \Delta Z_0) \label{eq:discHeight}
\end{equation}

\noindent
where $\rho^*_0$ and $\Delta Z_0$ are the stellar densities (in number of stars per unit of volume) at the plane and the disc scale height (in kpc), respectively. More precisely, $\Delta Z_0$ does not coincide with the scale-height of an exponential disc $h$, but is approximately twice, i.e. $\Delta Z_0 \approx 2h$ \citet[][]{Gilmore1990}. Equation~\ref{eq:discHeight} was obtained  solving the Poisson and Liouville equations under the assumption that the vertical distribution of the disc light is that of a locally isothermal sheet. 
A fit to the data using Eq.~\ref{eq:discHeight} is shown in Fig.~\ref{fig:scaleHeight} (blue line) providing a scale height of the LMC disc of $\Delta Z_0$=1.94$\pm$0.02 kpc.
However, since for an isothermal disc the value of $\Delta Z_0$ can depend on the radius, we have also fitted Eq.~\ref{eq:discHeight} in different annuli, ranging from 0.0 to 9.0 kpc.
Caused by the small number of DCEPs beyond 3.5--4 kpc from the LMC centre, that makes the fit of the data highly unstable, the last annulus is significantly larger than the others. The pulsators within the smallest radius are in large majority located on the bar, while the largest radius allows us to include almost all the DCEPs in our sample. The result of this exercise is reported in Table~\ref{tab:diskHeight}. The table shows a steady increase of the  $\Delta Z_0$ value towards the more external regions of the LMC. This result is in qualitative agreement with the work by \citet{Alves2000}, based on a study of carbon stars in the LMC. Studying the velocity dispersion of these stars, and assuming a vitalised disc, \citet{Alves2000} found a scale height that varies by a factor 5--6  (depending on the models) from $R=0.5$ kpc to $R=5.6$ kpc ("flaring" of the disc).

If we compare the scale height of the LMC disc with the scale height of the MW thin and thick discs, whose values at the solar location are $h_{\rm thin}$=300$\pm$60 pc and $h_{\rm thick}$=900$\pm$180 pc \citep[][]{Juric2008,Bland2016}, with the LMC value $h_{\rm LMC}\approx$975 pc,\footnote{As the MW scale heights were obtained with an exponential density law, we halved the value of $\Delta Z_0$=1.94 kpc obtained for the LMC disc scale height with the sech$^2$ law} it results that the scale height of the LMC disc is similar to that of the MW thick disc. 

We may wonder what the origin of this thick and flared disc is. According to earlier $N$-body simulations \citep[e.g.][]{Weinberg2000,Bekki2005}, the thick disc of the LMC is the result of the strong MW tidal field that heats the LMC disc stars at any orbit pericentre with respect to the Galaxy. The thickening of the LMC disc starts from the outer, more fragile part of the disc, where they prevail over the LMC’s self-gravity, producing the observed flaring \citep[][]{Weinberg2000,vanderMarel2002}. 

However, both \citet[][]{Weinberg2000} and \citet{Bekki2005} models used a 5--10 times smaller total mass for the LMC than what currently accepted \citep[e.g. $\sim 1.4 \times 10^{11}$ M$_{\odot}$][]{Cullinane2021}, so that the tidal forces of the MW are over-estimated. Moreover, we know now that the MCs have likely just passed the pericentre on their first infall into the MW potential \citep[][]{Kallivayalil2013}. Therefore there was not enough time to heat sufficiently the LMC disc with the MW tidal field.
Nevertheless, also the SMC can contribute to heat the LMC disc through its tidal forces. Indeed, even if at its present location the tidal strength applied by the SMC is about 17 times smaller than that of the MW \citep[][calculated adopting a mass of $\sim 3 \times 10^9$ M$_{\odot}$ for the SMC and a MW mass enclosed within the LMC orbit of $\sim 5 \times 10^{11}$ M$_{\odot}$]{vanderMarel2001}, we know from up to date $N$-body simulations that the SMC could have had a close encounter or a direct impact with the LMC (depending on the models) some 100--150 Myr ago \citep[e.g.][]{Zivick2018,Zivick2019} or a pericentric passage and a LMC disc plane crossing about 150 and 400 Myr ago, respectively \citep[][]{Cullinane2021}. In any case if the SMC approaches to less than 10 kpc of the LMC centre its tidal forces become predominant with respect to the MW's \citep{vanderMarel2002}, so that it is likely that the SMC had a prominent role in heating the LMC disc. 

\begin{table}
\caption{Scale-height values of the LMC disc calculated at different annuli from the galaxy centre. R1 and R2 are the initial and final radius adopted in the calculation; $\Delta Z_0$ is the scale height from Eq.~\ref{eq:discHeight} including the formal error from the fitting procedure; $\sigma$ is the dispersion of the residuals (data-fitted model) expressed in percentual; $N_*$ is the number of DCEPs used in the calculation.}
\footnotesize\setlength{\tabcolsep}{3pt}
\label{tab:diskHeight}
\begin{center}
\begin{tabular}{ccccccc}
\hline 
\noalign{\smallskip} 
R1 & R2 & $\rho^*_0$ &$\Delta Z_0$ & $\sigma$ & $N_*$  \\
(kpc) &   (kpc) & ($N_*$/kpc$^3$)  &  (kpc) &   per cent &  &  \\
\noalign{\smallskip} 
\noalign{\smallskip} 
\hline 
0 & 2.0   & 40.7$\pm$2.2 & 1.23$\pm$0.08 & 7.1 & 1199 \\ 
1 & 3.0   & 29.8$\pm$0.8 & 1.68$\pm$0.06 & 4.6 & 2411 \\
2 & 4.0   & 15.1$\pm$0.6 & 2.21$\pm$0.11 & 7.4 & 2388 \\
3 & 6.0   & 2.8$\pm$0.1 & 3.29$\pm$0.18 & 8.8 & 1482 \\
3.5 & 9.0 & 0.6$\pm$0.1 & 4.43$\pm$0.28 & 10.1 & 1001 \\
\noalign{\smallskip}
\hline
\noalign{\smallskip}
0 & 9.0 & 4.20$\pm$0.04 &1.94$\pm$0.02 & 1.5 & 4220 \\
\noalign{\smallskip}
\hline
\end{tabular}
\end{center}
\end{table}


It is worth noting that tidal forces are not the only possible cause for the thickening of the LMC disc. For instance, to explain the presence of a counter-rotating population in the disc of the LMC found empirically by \citet[][]{Olsen2011}, \citet{Armstrong2018} hypothesised that it could have been produced by a minor merging of a satellite galaxy with the LMC. Their hydrodynamical simulations reveal that a retrograde merging of a dwarf galaxy with mass $\sim 5 \times 10^{10}$ M$_{\odot}$ (about 20 times smaller than the LMC) occurred 3--5 Gyr ago can reproduce the counter-rotating population and would also have the side effect of thickening the LMC disk. Although \citet{Armstrong2018} do not provide a quantitative estimate of the thickening of the LMC disc produced by merging, there is an additional channel able to produce the huge thickness of the LMC disc. 
Interestingly, the hypothesis that the LMC experienced at least one minor merger in the past was nicely confirmed in a recent work by \citet[][]{Mucciarelli2021} who found that the old LMC globular cluster (GC) NGC\,2005 has a chemical composition not compatible with that of the other old LMC's GCs\footnote{It is important to note that \citet[][]{Mucciarelli2021} analysed the data of all the LMC GCs in a homogeneous way}. \citet[][]{Mucciarelli2021} argue that NGC\,2005 originated in a galaxy
where star formation was much less efficient than in the LMC, similar to the dwarf spheroidal galaxies (dSph) currently orbiting the MW having masses of the order of  $1-100 \times 10^6$ M$_{\odot}$. 
Even if these masses are much smaller than that of the dwarf galaxy hypothesised by \citet{Armstrong2018}, none the less, the \citet[][]{Mucciarelli2021} results demonstrate that the merging channel is not a remote conjecture to explain the thickness of the LMC disc. 

To conclude, the observational results presented here give new constraints on the last LMC--SMC--MW interaction, and new modelling is needed to assess in a more quantitatively way the effectiveness of tidal interaction or merging channels to reproduce the observed properties of the LMC disc as revealed by DCEPs.

\section{Kinematics of the disc}

The kinematics of the disc can be then studied by extracting the astrometry of the sources from the first instalment of the Third Data Release of the {\it Gaia} catalogue \citep[EDR3][]{Brown2021}.
In this section, we thus model the kinematics of the disc of young stars, and compare with results presented by \citet{Luri2021} for other stellar evolutionary phases. To achieve this goal, we used the tools developed by \citet{Helmi2018} and \citet{Luri2021}. As the line-of-sight velocities of the sources are not known, it is not possible to constrain the 3D velocity space for them. For this reason, the modelling  assumes that in a cylindrical frame, the vertical component cancels out, and the {\it Gaia} proper motions can be used to build maps of radial and tangential velocities for both the ordered and random motions.

The maps were constructed by fixing the parameters of the disc to the kinematic values measured by \citet{Luri2021}, corresponding to a position angle of the major axis of $\Omega = 310$ degr, an inclination of $i = 34$ deg, and a centre-of-mass position and proper motion of $\alpha_0 = 81.28$ deg, $\delta_0 = -69.78$ deg, $\mu_{x,0} = 1.858$ \masyr, $\mu_{y,0} = 0.385$ \masyr\ and $\mu_{z,0} = 1.104$ \masyr. We also considered that these parameters do not vary as a function of radius. Because these parameters were derived from $11,156,431$ stars of many evolutionary phases, the DCEPs representing only less than 2 per cent of the Blue Loop sub-sample and 0.04 per cent of the total LMC sample from \citet{Luri2021}, it is not surprising to see that the kinematic orientation differs from the one inferred with the morphology of the DCEPs disc. In particular, the morphologically-based inclination finds a disc which is more face-on by $9$ deg than found by the kinematics, and with a major axis  position angle differing by $\sim 10$ deg. As a consequence, the galactocentric radius defined in this Section is that of a kinematic frame which differs slightly from the one used in previous sections. To avoid confusion with previous sections, we refer to the radius as \rkin\ hereafter. We then constructed maps of \vrad, \vtan, \srad\ and \stan, each one being an image of $64\times64$ squared pixels, adopting a pixel size of $0.25$ deg ($\sim 220$ pc at the distance of the LMC), thus corresponding to a sampling $6$ times coarser than in \citet{Luri2021}, but sufficient for our purpose.  

\begin{figure*}
 \includegraphics[width=0.5\textwidth]{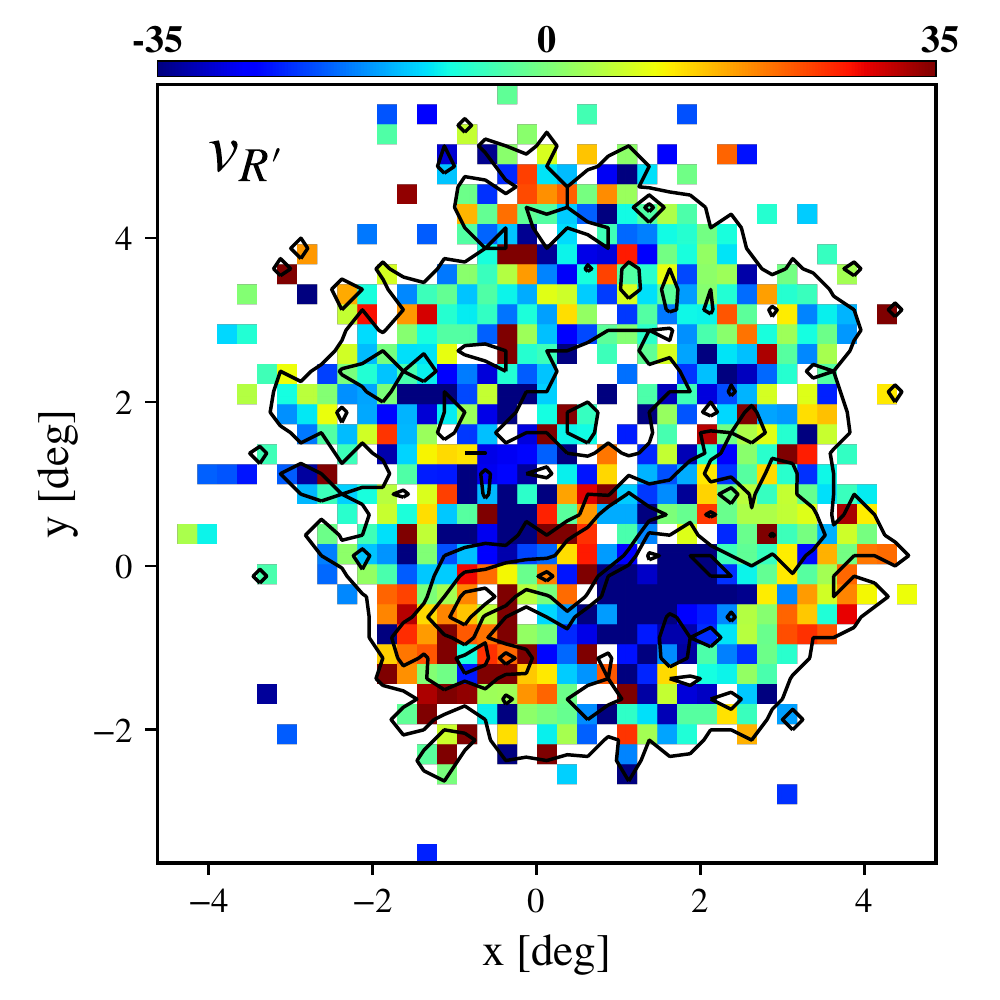}\includegraphics[width=0.5\textwidth]{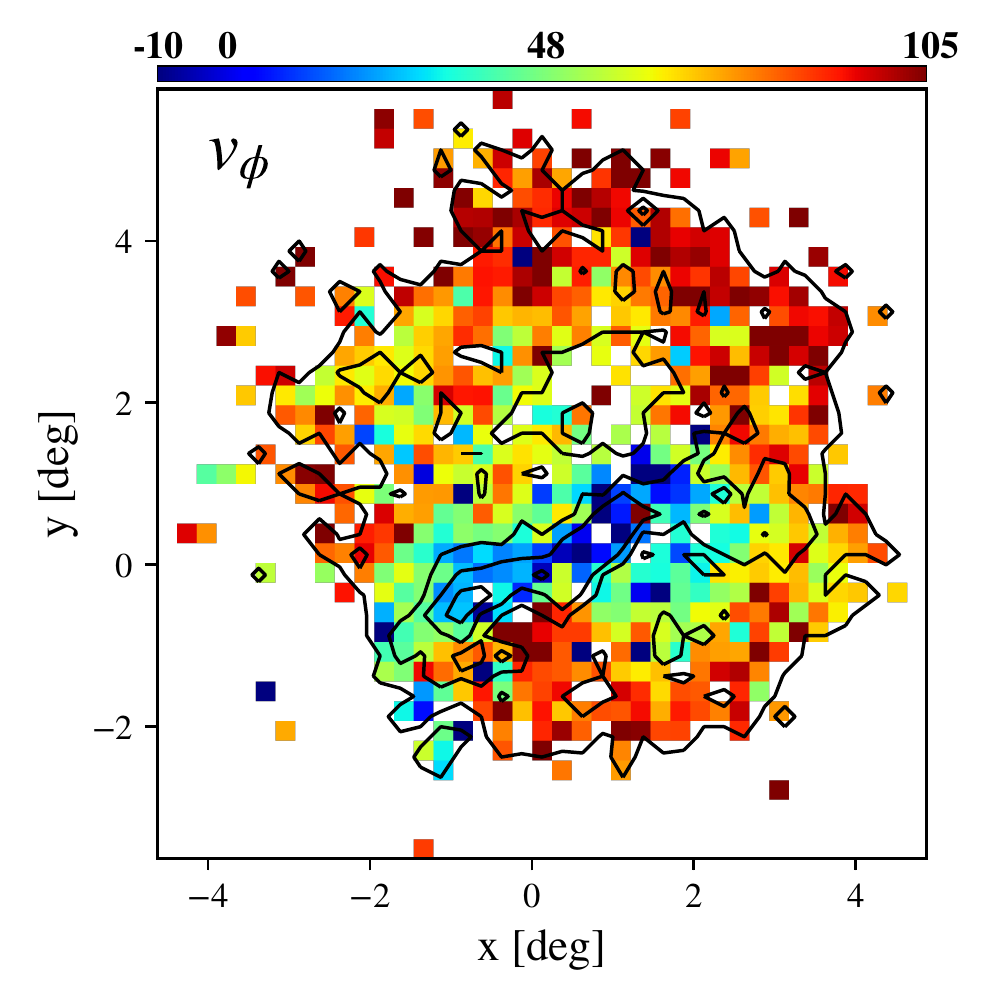}
 \includegraphics[width=0.5\textwidth]{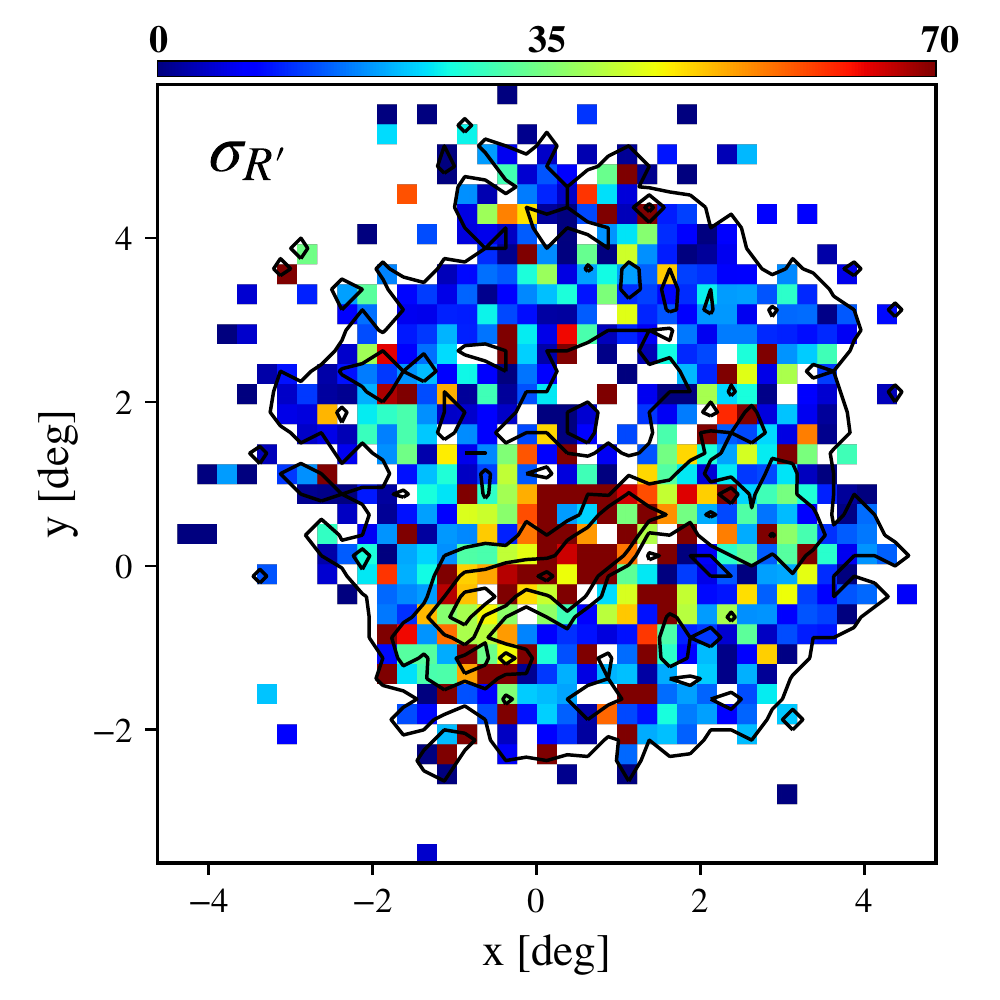}\includegraphics[width=0.5\textwidth]{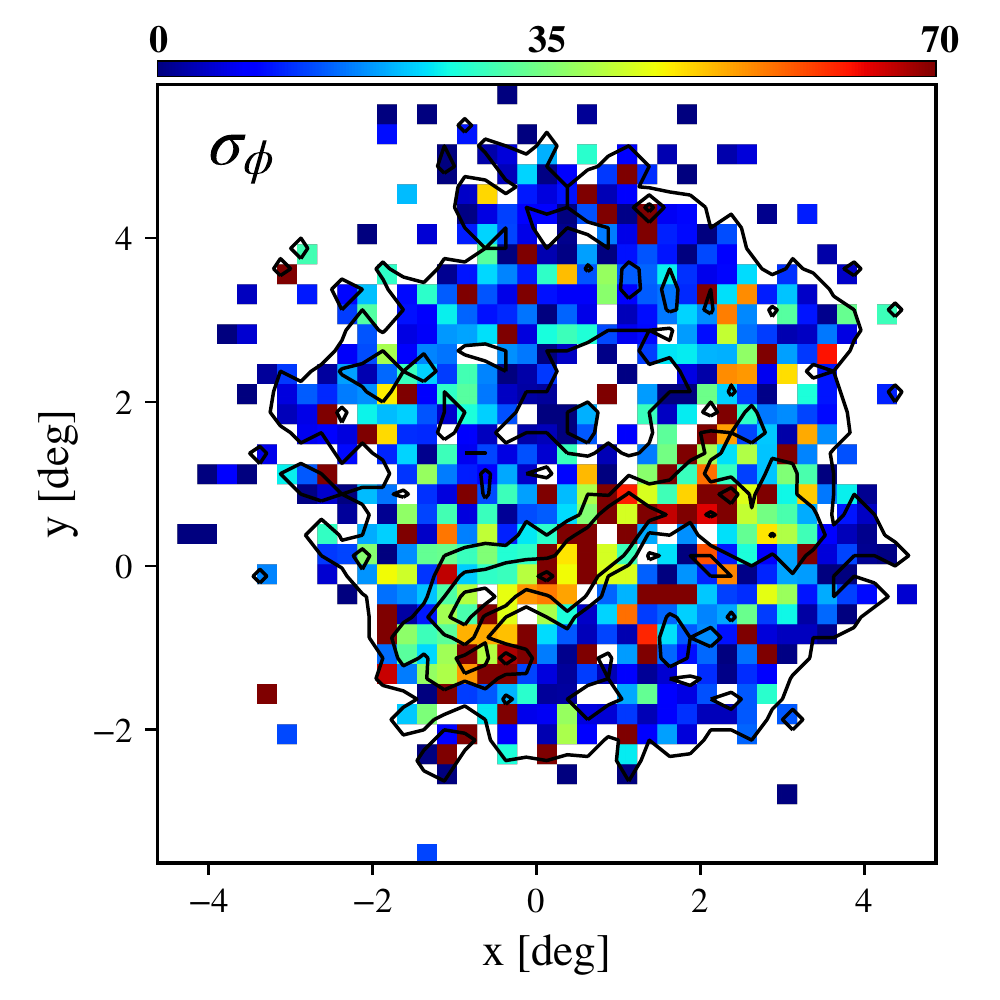}
\caption{Velocity maps of the LMC disc of DCEPs. The ordered and random motions are shown in the upper and bottom rows, respectively, for the radial and tangential components (left and right columns, respectively). The unit is \kms. The velocity range is shown as a top colour-bar in each panel, and has been chosen to highlight more easily the variations in the bar and spiral arms. The 4 contours represent the stellar density (2,10, 20 and 50 stars per squared pixel of 0.25 deg).}
 \label{fig:velomaps}
\end{figure*}

Figure~\ref{fig:velomaps} shows the four velocity maps on which contours of stellar density have been superimposed. The radial motion of DCEPs is mainly negative (top-left panel). This suggests that the orbits of stars around the galactic centre are principally contracting, i.e. that stars are losing angular momentum. 
A strong variation of \vrad\ is observed within the bar region. The ordered radial velocity is larger along the bar at higher stellar densities (e.g. $30 < v_{R'} < 90$ \kms\ around $(x,y) = (-1,~-1)$), and  smaller in two  regions oriented almost perpendicularly to the bar (e.g. \vrad\ down to $-80$ \kms\ at $(x,y) \sim (1.5,-1)$, within the SWS structure). The important \vrad\ streaming  corresponds to radial motions globally directed outwards along the bar, and inwards nearly perpendicularly to the bar around the centre. 

The rotation velocity (top-right panel) increases continuously with radius, and shows streaming motions in the bar region. The region of lowest tangential velocity is beautifully observed aligned with the bar, in which pixels with negative \vtan\ are seen. This is not caused by noise, as the density of stars in those pixels are among the highest in the map (at least 10 stars per pixel).  \citet{Luri2021} already reported this finding for all their stellar samples. There is thus a hint of counter-rotation in the stellar bar of the LMC. Moreover,  the location $(x,y) \sim (1.5,-1)$ kpc is particularly interesting because stars are moving faster inwards than rotating around the galactic centre ($v_{R'} > v_\phi$). Finally, the \vtan\ map shows streaming motions along the spiral arm to the North, in which the velocity decreases towards the East.

The bottom panels of Fig.~\ref{fig:velomaps} show a larger velocity dispersion all along the stellar bar for both the radial and tangential  components,  as well as in the two regions of lowest \vrad\ roughly perpendicular to the bar, but for \srad\ only.  The radial random motion is mainly larger than the tangential component.

It is worth noting here that \citet{Bovy2019} used APOGEE (Apache Point Observatory Galactic Evolution Experiment) Data Release 16 spectroscopy \citep{Majewski2017}, cross-matched with {\it Gaia} DR2 astrometry \citep{Brown2018}, to construct partial stellar radial and tangential velocity maps in the direction of the bulge and the bar of the Galaxy. They identified a pattern of negative--positive \vrad\ across the galactic bar, also showing lower azimuthal velocity along the bar. Their result was supported by a numerical $N$-body simulation of a barred, MW-like disc \citep[see Fig. 3 of \citet{Bovy2019}, adapted from a simulation by][]{Kawata2017}. 
The significant variations of \vrad\ and \vtan\ found inside and across the LMC bar are thus reminiscent of those predicted by the numerical simulation. This finding is even more striking in the LMC than the MW since the entirety of the bar region is covered by our data, and is observed in many other kinematic stellar tracers as well  \citep[see fig. B.2 in][]{Luri2021}. A difference  however exists in the orientation of the LMC quadrupole \vrad\ pattern, which puts the inwards motion (outwards, respectively)  almost aligned (perpendicular) to the young DCEP bar, while it is not the case with the Galactic bar and the simulation. We postpone to another work the detailed comparison of the ``four-leaf clover''  shape for \vrad\ in the various LMC stellar populations  with numerical models.

\begin{figure*}
 \includegraphics[width=0.5\textwidth]{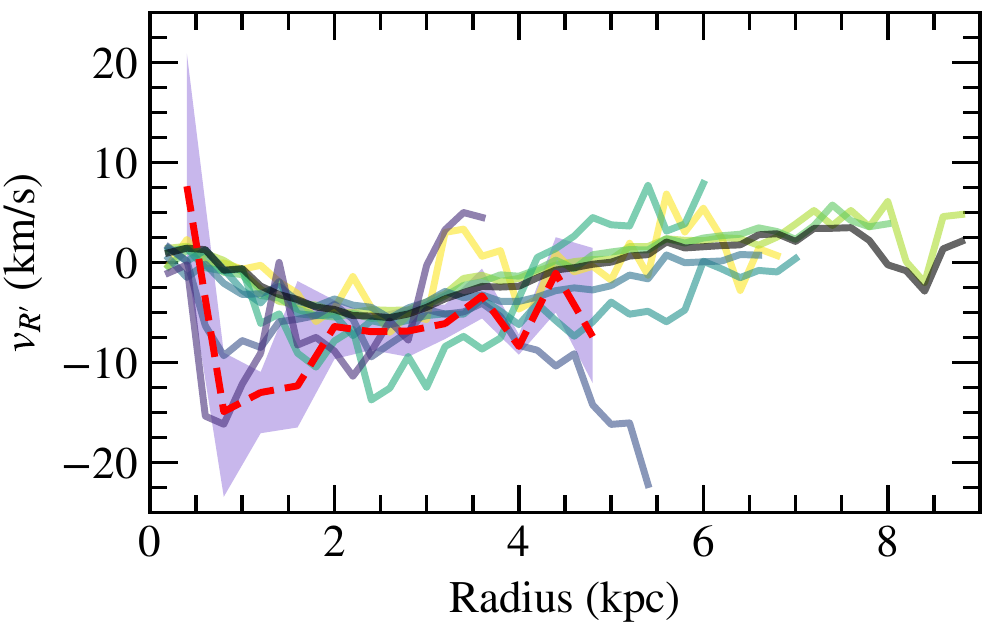}\includegraphics[width=0.5\textwidth]{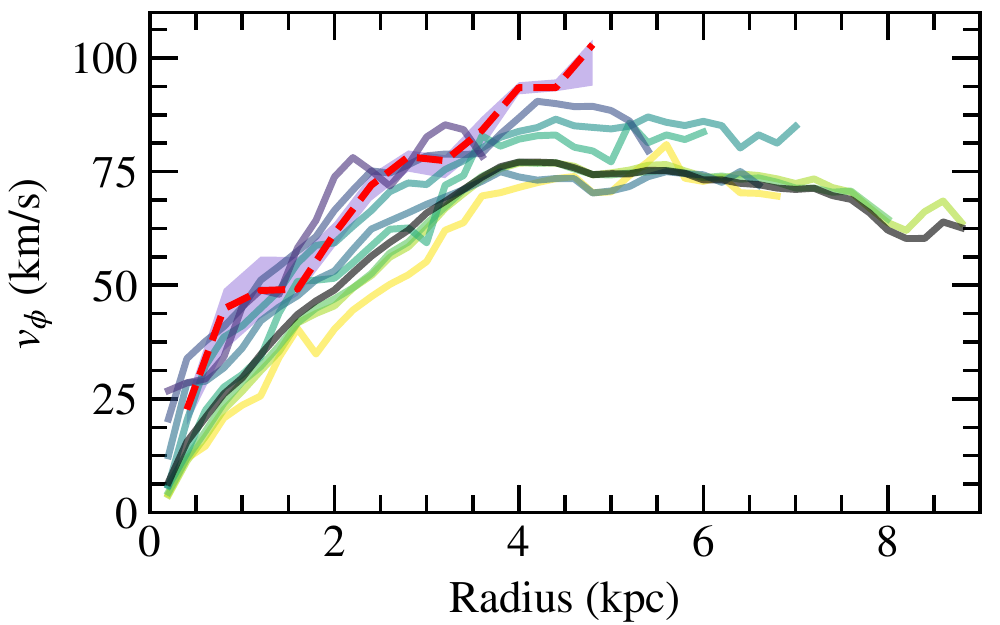}
 \includegraphics[width=0.5\textwidth]{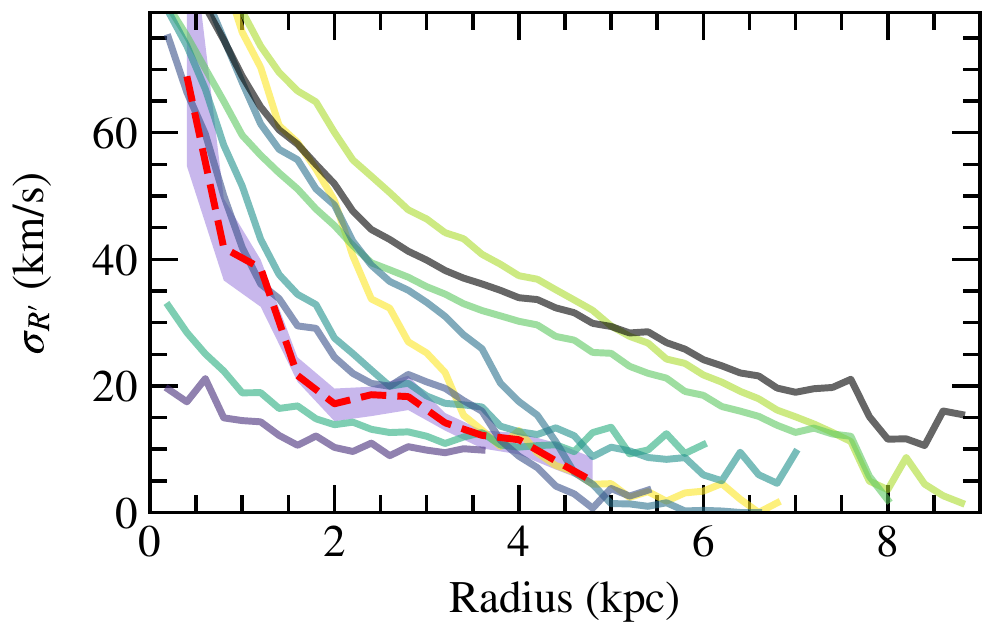}\includegraphics[width=0.5\textwidth]{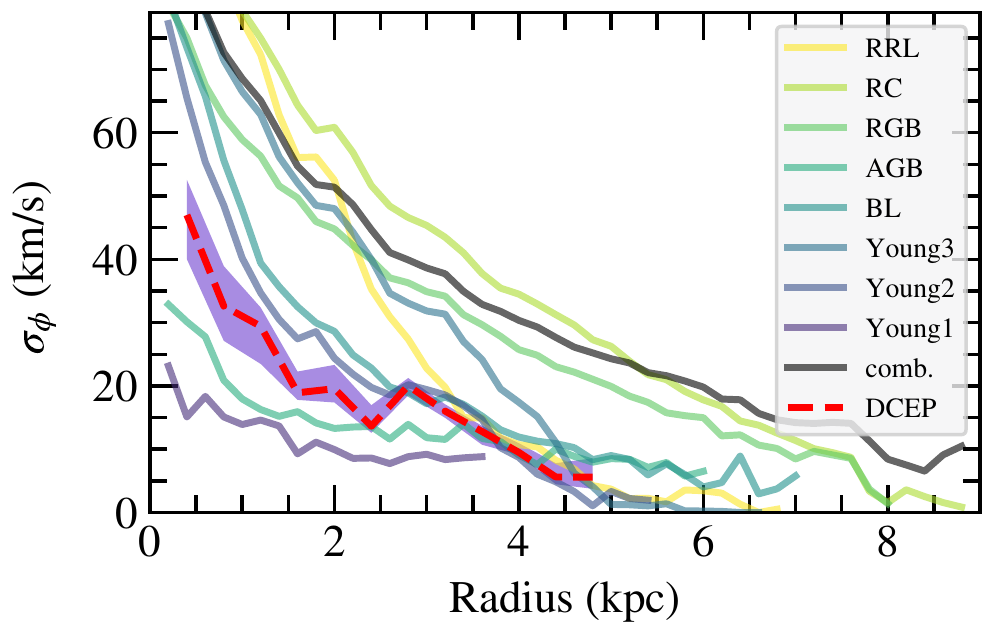}
\caption{Velocity curves of the LMC disc of DCEPs. The ordered and random motions are shown in the upper and bottom rows, respectively, for the radial and tangential components (left and right columns, respectively). The DCEPs curves are shown as dashed red lines.  Other lines are for the various  stellar evolutionary phases studied in \citet{Luri2021}, as listed in the bottom right panel. They correspond to RR Lyrae (`RRL'), Red Clump (`RC'), Red Giant Branch (`RGB'), Asymptotic Giant Branch (`AGB'), Blue Loop (`BL'), Main Sequence stars (`Young 1, 2, 3') and a sample combining the 8 subsamples (`comb.').}
\label{fig:velocurves}
\end{figure*}

We also inferred the velocity profiles from the velocity fields, similarly as in \citet{Luri2021}. The velocity is the median value of pixels from the maps located inside a bin centred on a given galactocentric radius. The width of the radial bin is 400 pc, yielding 15 stars in the innermost bin at $R' = 0.4$ kpc, and 215 in the outermost bin at $R'=4.8$ kpc. Bootstrap resamplings were performed to constrain the velocity uncertainties, which were measured at the 16th and 84th percentiles of the generated velocity distributions. The velocity profiles of the DCEPs are shown in Fig.~\ref{fig:velocurves} as red dashed lines, and the uncertainties as shaded areas. 

Figure ~\ref{fig:velocurves} is particularly interesting in the comparison with the curves of other stellar populations. The amplitude of the tangential velocity of the DCEPs are among the  largest among all stellar evolutionary phases. Compared to the RC sample of \citet{Luri2021}, differences of up to $\sim 30$ \kms\ are observed. This is not surprising as younger stars are expected to exhibit larger rotation motions than older ones, as they rotate at velocities closer to the circular velocity because of a lower asymmetric drift exerted on them.  Moreover, no velocity plateau or decline in the DCEPs rotation curve is observed such as those seen in every other kinematic stellar tracers, as a result from the continuous rise in the tangential velocity field mentioned above. 

As for the \vrad\ profile, it is among the lowest observed. This can be explained such that DCEPs were born more recently from gas having likely significant radial motion, and therefore their disc has not yet had time to relax, contrary to more evolved stars.  Note also that the dip of \vrad\ does not occur at the same radius of the minimum measured in the disc(s) of more evolved stars. The sharp fall and larger uncertainty of \vrad\ within the first kpc are caused by the significant velocity streaming discussed above. 

Another consequence of the strong  variation of \vrad\ at low radius is the steep inner slope for the \srad\ profile. We also see here that in the inner kpc, \srad\  is by 10 to 20 \kms\ larger than \stan, while beyond that radius \srad\ can be sometimes smaller or larger than \stan, but to a weaker extent. The curves and maps of random motions illustrate nicely an anisotropic velocity ellipsoid for the LMC disc of DCEPs in the bar region. It is indeed clear that the stellar orbits  are radially biased for $R' < 2$ kpc.  We measure a planar velocity anisotropy parameter, defined as $\beta_{\phi, R'} = 1 - \sigma^2_\phi/\sigma^2_{R'}$, within 0 and 0.5. This contrasts with the rather isotropic velocity ellipsoid that we observe for the other stellar phases ($\beta_{\phi, R'} \sim 0$ in the bar region, on average). A more complete analysis of the shape of the stellar velocity ellipsoid is however beyond the scope of this article. 

\section{Discussion and conclusions}

We presented the light-curves of 4408 DCEPs observed by the VMC survey in the LMC over an area of more than 100 deg$^2$. 
We provided $Y$, $J$, and $K_\mathrm{s}$ average magnitudes, amplitudes and
relative errors for these DCEPs, calculated based on an extended set of templates derived from our own data. The errors were estimated using the bootstrap technique.  With respect to similar investigations using NIR magnitudes our sample is more: i)  homogeneous, as 98\% of the data come from a unique telescope/instrument \citep[i.e. apart from the 92 objects with photometry by][]{Persson2004}; ii) accurate, as our light curves are or well-sampled in $K_\mathrm{s}$ and moderately well-sampled with at least four epoch data in  $J$ and $Y$; iii) complete, as it includes about 97 per cent of the known DCEPs in the LMC. In particular, we publish for the first time the NIR photometry of the faint, short-period, 1O-mode pulsators.  

The intensity-averaged magnitudes in the VISTA $Y$, $J$, and
$K_\mathrm{s}$ filters were complemented by optical $V$-band
data and periods from the literature to construct multi-filter $PL$/$PW$
relations for LMC DCEPs. The $PL$ and 
$PW$ relations in the $V, J$, and $K_\mathrm{s}$ bands for  
F- and 1O-mode LMC DCEPs presented here are the most precise to date.

During the process of derivation of the $PL$/$PW$ relations we discovered for the first time a break for the 1O-mode pulsators at $P$=0.58 d. We explain this feature by observing that the IS is fainter and narrower for such short period DCEPs which are supposed to be in their first crossing of the IS.    

We adopted updated period--age--metallicity relations to calculate the ages of the DCEPs in the LMC. We find an average age of about 100 Mys with a dispersion of 33 Myr. However, looking at the spatial distribution of DCEPs in different intervals of age, we discovered that the younger objects have a roundish distribution approximately corresponding to the bar/spiral arm structure, but with fine sub-structuring. On the other hand, the older objects (mainly short period 1O-mode pulsators) have a more diffuse distribution. 

The $PL$/$PW$ relations calculated here have been used to estimate individual distances for the DCEPs in the LMC, with relative individual distance precision of better than 1 kpc in the best case. These data allowed us to calculate the viewing angles of the LMC using a larger sample of DCEPs than previous similar investigations. We found: $\theta$=145.6$\pm$1.0 deg and $i$=25.7$\pm$0.4 deg. These values are in agreement with the recent literature for what concerns the inclination $i$, while are in disagreement concerning the value of $\theta$. We explain this occurrence with the larger and more extended sample adopted in our work. 
As found in previous works, we find that the bar and the disc of the LMC have different viewing angles.

The high precision of our relative distances allowed us to depict the most accurate 3D distribution of the DCEPs in the LMC to date. We found that the bar and the disc of the LMC show several sub-structures, confirming the general structuring described e.g. by \citet{Jacy2016} but adding new finer details. The bar seems to be composed of two clearly distinct components, the EB and the WB, separated on average by 1--2 kpc, with the EB being more detached from the disc than the WB. The disc appears to be warped in the north-western direction by about 1 kpc. The spiral arms also show a complex structure. In particular, the NA1 is formed by an eastern and a western component. The eastern component is closer to us and approximately aligned with the EB and SES, while the western component is aligned with the NA2 and the WB. The apparent association of these structures can be noticed also in the ages of their formation. Indeed, Fig.~\ref{fig:mapAges} shows that the EB and the eastern part of the NA1 continued to  form DCEPs up to about 30 Myr ago and in particular in the interval 30--60 Myr, when the star formation in the WB, the western part of NA1 and the NA2 decreased significantly. 
We can speculate that there is an association between spatial distribution and star formation history, hypothesising that in different parts of the LMC the availability of gas and dust for the star formation was different. However, the causes of this difference should still be investigated.

The overall distribution of LMC DCEP ages was modelled using two Gaussian functions with means 93 and 159 Myr and dispersion 21 and
17 Myr, respectively. We roughly assumed twice the dispersion, i.e.
$\sim$40 Myr, as the time-scale of both star formation episodes that
produced DCEPs in the LMC. Therefore, we can speculate that 
the DCEPs in the LMC were formed in two main episodes
of star formation lasting $\sim$40 Myr which happened 93 and 159
Myr ago. The first event was significantly less efficient than the second, that
 produced more than 80 per cent of the DCEPs in the LMC.

We calculated the disc height of the LMC finding that, apart from the central regions, the density profile of the LMC DCEPs follows the expected sech$^2$ distribution. We do not find signs of flaring of the disc. The inferred scale height of the LMC disc is 1.94 kpc, corresponding to 0.97 kpc for an exponential disc. This value is similar to that of the Galactic thick disc. We discuss different physical mechanisms able to thicken the LMC disc, including tidal interactions and merging. However, only up to date $N$-body simulations can allow us to understand which phenomenon is more likely responsible for the thickening of the LMC disc. 

The cross-match with astrometric data from {\it Gaia} EDR3 has allowed us to build maps and radial profiles for both the ordered and random motions of the radial and tangential velocities in the plane of the LMC. The kinematics is tightly correlated to the bar and spiral structures. In particular, a quadrupole pattern of inwards and outwards radial motion is observed in the LMC bar, also with lowest rotational motions well aligned with the bar. These signatures are very reminiscent of the kinematics of stars in the MW stellar bar, and expectations from simulated barred objects. The stellar velocity ellipsoid is anisotropic in the bar region, with stellar orbits radially oriented, and there are even locations where the radial velocity exceeds the tangential motion of stars. Overall, a bulk inwards motion is observed in the disc of DCEPs.  Future work is needed to assess the differences of orbital structure as a function of stellar age in the bar and spiral arms of the LMC. As for the rotation curve, it globally exceeds some of the curves presented in \citet{Luri2021} for other stellar populations, as a likely consequence of lower asymmetric drift in the disc of DCEPs than for more evolved stars.

The geometric, kinematic and evolutionary properties of the DCEPs in the LMC discussed in this work provide an important set of constraints for future $N$-body simulations with the final aim of reconstructing the recent history of the interaction between LMC, SMC and MW.   
In this context, an unprecedented contribution is expected in the next years by the massive spectroscopic surveys of pulsating (and non-pulsating) stars that will be carried out with instruments such as 4MOST \citep[4-metre Multi-Object Spectroscopic Telescope for VISTA;][]{deJong2019} and MOONS \citep[Multi Object Optical and Near-infrared Spectrograph for the Very Large Telescope;][]{Cirasuolo2020}. These data, in conjunction with the more precise PM expected from the {\it Gaia} Data Release 4 will eventually allow us to disclose the ``true'' story of the formation and evolution of the LMC and of the Magellanic System as a whole.

\section*{Acknowledgements}
We thank our anonymous Referee for their constructive and helpful comments.
This research has made use of the SIMBAD database, operated at CDS, Strasbourg, France.
This research was supported in part by the Australian Research Council Centre of Excellence for All Sky Astrophysics in 3 Dimensions (ASTRO 3D), through project number CE170100013.

\section*{Data Availability}

All the data used in this work are described in the tables above which will be published in their entirety in electronic form as part of this paper.  




\bibliographystyle{mnras}







\bsp	
\label{lastpage}
\end{document}